\documentclass[tikz,11pt]{article}
\pdfoutput=1

\usepackage[utf8]{inputenc}
\usepackage{multirow}
\usepackage{amsmath, amsfonts, amssymb}
\usepackage{comment}
\usepackage{graphicx}
\usepackage{psfrag}
\usepackage{amsthm}
\usepackage[usenames,dvipsnames,svgnames,table]{xcolor}
\usepackage{enumerate}
\usepackage{arydshln}
\usepackage{soul}
\usepackage{slashed}
\usepackage{subfig}
\usepackage{mathrsfs}
\usepackage{a4wide}
\usepackage{tikz}
\usepackage{tikz-cd}
\usetikzlibrary{shapes.geometric}
\usepackage{tcolorbox}
\usepackage{mathtools}
\usepackage{xfrac}
\usepackage{subcaption}
\usepackage{amsmath}
\usepackage{xcolor}
\usetikzlibrary{matrix,calc}

\usepackage{epsf}

\usepackage{color}
\definecolor{dark-gray}{gray}{0.20}
\definecolor{gray}{gray}{0.30}
\definecolor{light-gray}{gray}{0.80}
\definecolor{dark-red}{rgb}{0.7,0,0}
\definecolor{dark-green}{rgb}{0.1,0.4,0}
\definecolor{dark-blue}{rgb}{0.3,0.3,0.7}
\definecolor{light-blue}{rgb}{0.8,0.8,1}
\definecolor{swamp}{RGB}{240, 199, 197}
\definecolor{landscape}{RGB}{180, 250, 199}
\definecolor{undecided}{RGB}{252, 252, 197}

\usepackage{pifont}

\usepackage{setspace}

\newcommand{\beq}{\begin{equation}}  \newcommand{\eeq}{\end{equation}}
\newcommand{\bal}{\begin{aligned}}   \newcommand{\eal}{\end{aligned}}
\newcommand{\be}{\begin{equation}}
\newcommand{\ee}{\end{equation}}
\newcommand{\ba}{\begin{aligned}}   \newcommand{\ea}{\end{aligned}}

\def\beqa{\begin{eqnarray}}
\def\eeqa{\end{eqnarray}}

\def\be{\begin{equation}}
\def\ee{\end{equation}}
\def\bea{\begin{eqnarray}}
\def\eea{\end{eqnarray}}

\def\simleq{\; \raise0.3ex\hbox{$<$\kern-0.75em
      \raise-1.1ex\hbox{$\sim$}}\; }
   \def\simgeq{\; \raise0.3ex\hbox{$>$\kern-0.75em
      \raise-1.1ex\hbox{$\sim$}}\; }

\numberwithin{equation}{section}

\usepackage{jheppub}
\usepackage{hyperref}
\usepackage{cleveref}

\hypersetup{
	colorlinks=true,
	linkcolor=dark-blue,
	citecolor=dark-red,
	urlcolor=dark-green,
	linktoc=page
}

\theoremstyle{remark}

\crefname{appendix}{Appendix}{Appendices}

\title{\centering The CFT Distance Conjecture and Tensionless String Limits in $\mathcal N=2$ Quiver Gauge Theories}

\author{Jos\'e Calder\'on-Infante$^1$}
\author{and Amineh Mohseni$^{2}$}
\affiliation{$^1$Walter Burke Institute for Theoretical Physics, \\ California Institute of Technology, Pasadena, CA 91125, USA}
\affiliation{$^{2}$Jefferson Physical Laboratory, Harvard University, Cambridge, MA 02138, USA}

\emailAdd{joseci@caltech.edu, amohseni@g.harvard.edu}

\abstract{We initiate the study of infinite-distance limits on (complex) multi-dimensional conformal manifolds of 4d SCFTs and their bulk interpretation as tensionless-string limits in AdS/CFT. In particular, we focus on 4d $\mathcal{N}=2$ $SU$ quiver gauge theories with hypermultiplets in the bifundamental and fundamental representations. In the overall-free limit, we compute the large-$N$ Hagedorn temperature $T_H$, which governs the stringy exponential growth of the density of states at high energies. We argue that this quantity determines the type of stringy ultraviolet completion in the bulk: it captures the type of string theory in which the bulk physics is embedded while remaining insensitive to detailed geometric data. For linear quivers, we find that $T_H$ depends only on the quiver length, which is tied to the number of NS5-branes in the underlying brane construction and, in turn, to the string theory in which the bulk is embedded. For holographic quivers, where we impose that the two central charges $a$ and $c$ coincide in the large-$N$ limit, we show that $T_H$ coincides with that of $\mathcal{N}=4$ SYM, which befits the 10d Type IIB description of their gravitational duals. We also analyze the exponential rate $\alpha$, which controls how the leading tower of higher-spin currents becomes conserved in these limits, as suggested by the CFT Distance Conjecture. In the large-$N$ regime, we derive sharp bounds on the minimal rate, $1/\sqrt{2}\le \alpha_{\min}\le \sqrt{2/3}$, attained in the overall-free limit. Moreover, we prove that the universal lower bound $\alpha\ge 1/\sqrt{2}$ holds, including at finite $N$. Finally, we go beyond the overall-free ray by characterizing the convex hull of the $\vec{\alpha}$-vectors that encode the exponential rate of the higher-spin towers along any (partial) weak-coupling limit.
}

\begin{document}
\hypersetup{pageanchor=false}
\makeatletter
\let\old@fpheader\@fpheader\
\renewcommand{\@fpheader}{  \vspace*{-0.1cm} \hfill CALT-TH 2026-003}

\makeatother

\maketitle

\newpage

\section{Introduction}\label{sec:intro}
The Swampland program in asymptotically Anti-de Sitter (AdS) spacetimes and the conformal bootstrap program are closely connected through AdS/CFT. The Swampland program seeks to establish universal consistency conditions that any theory of quantum gravity must satisfy \cite{Vafa:2005ui}. On the other hand, the conformal bootstrap constrains conformal field theories (CFTs) from first principles such as symmetry, unitarity, and causality. Via AdS/CFT, any local,\footnote{Locality implies the existence of a conserved, traceless stress tensor. Through the AdS/CFT dictionary, this operator is dual to the graviton in the bulk.} unitary CFT defines a consistent quantum gravity theory in AdS. When the bulk theory is well-described by Einstein gravity at low-energies, the CFT is said to be \emph{holographic}. These CFTs satisfy extra conditions such as large central charge, large higher-spin gap, and sparseness of the low-spin spectrum (see e.g. \cite{Heemskerk:2009pn,El-Showk:2011yvt,Alday:2019qrf}). This correspondence between theories of quantum gravity in AdS and CFTs makes it natural to investigate Swampland criteria from a CFT viewpoint, where conformal bootstrap methods can provide robust constraints. A particularly fruitful arena for this \emph{holographic Swampland} approach is the study of the Distance Conjecture \cite{Ooguri:2006in}, one of the central proposals in the Swampland program.

The Distance Conjecture states that approaching an infinite-distance limit in moduli space is accompanied by the emergence of an infinite tower of states whose masses decay exponentially with the distance. Various studies of infinite-distance limits in string compactifications have led to a classification of these limits according to the microscopic origin of the tower and the nature of the UV completion above its scale \cite{Grimm:2018ohb, Grimm:2018cpv, Corvilain:2018lgw, Lee:2018urn, Lee:2019xtm, Lee:2019wij,Marchesano:2019ifh,Baume:2019sry,Klaewer:2020lfg,Lanza:2021udy,Lee:2021qkx,Lee:2021usk,Etheredge:2022opl,Alvarez-Garcia:2023gdd,Alvarez-Garcia:2023qqj,Etheredge:2023odp,Etheredge:2023usk,Hassfeld:2025uoy,Grieco:2025bjy,Monnee:2025ynn,Monnee:2025msf}. Recent studies have also addressed the relationship between the tower of states and the quantum-gravity cutoff, as encoded in the species scale \cite{Dvali:2007hz,Dvali:2007wp, Dvali:2009ks,Dvali:2010vm}, their appearance in higher-curvature corrections, and their relation to black hole physics \cite{vandeHeisteeg:2022btw,Cribiori:2022nke,Cribiori:2023ffn,Cribiori:2023sch,Calderon-Infante:2023uhz,vandeHeisteeg:2023dlw, Castellano:2023stg, Castellano:2023jjt, Castellano:2023aum,Basile:2024dqq,Herraez:2024kux,ValeixoBento:2025bmv,Calderon-Infante:2025ldq,Herraez:2025clp,Aoufia:2025ppe}. This improved understanding has led to sharp bounds on the exponential decay rates of both the tower and the species scale \cite{Etheredge:2022opl, vandeHeisteeg:2023ubh, Calderon-Infante:2023ler}. Furthermore, these advances have motivated a classification of the possible structures of towers and their interplay \cite{Etheredge:2024tok}, as encoded in the convex-hull formulation for the Distance Conjecture proposed in \cite{Calderon-Infante:2020dhm}.\footnote{For analogous statements also involving extended objects, rather than only towers of particle states, see~\cite{Font:2019cxq, Etheredge:2024amg, Etheredge:2024tok}.} Recent efforts also include proposals for notions of distance beyond strict moduli spaces that also take into account the scalar potential \cite{Palti:2024voy, Mohseni:2024njl, Debusschere:2024rmi,Palti:2025ydz}.

In the context of AdS/CFT, where the moduli space of AdS vacua is mapped to the CFT conformal manifold, the Distance Conjecture was first explored in \cite{Baume:2020dqd, Perlmutter:2020buo}. It was found that, for 4d SCFTs, the Distance Conjecture in the bulk is naturally realized as (possibly partial) weak-coupling limits at infinite distance in the conformal manifold. This led to the proposal of the \emph{CFT Distance Conjecture} \cite{Perlmutter:2020buo}, which posits that, for local and unitary CFTs in $d>2$, infinite-distance limits in the conformal manifold are in one-to-one correspondence with limits in which a tower of higher-spin (HS) currents become conserved, with anomalous dimensions that approach zero exponentially fast with the Zamolodchikov distance. For four-dimensional theories, the authors of \cite{Perlmutter:2020buo} also proposed a sharp lower bound on the exponential rate of the HS tower in the bulk in Planck units: $\alpha \geq 1/\sqrt{2}$. Remarkably, that HS currents can become conserved in the conformal manifold only at infinite-distance points was proven in full generality in \cite{Baume:2023msm} using CFT techniques.\footnote{An analogous statement for two-dimensional CFTs was subsequently proven in \cite{Ooguri:2024ofs}, namely that the scalar gap of a 2d CFT can only go to zero in the conformal manifold at infinite-distance points. Furthermore, \cite{Ooguri:2024ofs} also shows that the scalar gap behaves exponentially with the Zamolodchikov distance in these limits and put sharp bounds on the exponential rate.} More recently, the bulk origin of the HS tower and the UV completion above its mass scale were investigated in \cite{Calderon-Infante:2024oed}, where it was proposed that these infinite-distance limits correspond to \emph{tensionless string limits} in AdS. To test this idea, it was shown that the weak-coupling infinite-distance limit of all four-dimensional superconformal gauge theories with simple gauge group and large $N$ limit fall into three universality classes that share (at large $N$) the same value of $\alpha$, the same ratio of conformal anomalies $a/c$, and the same Hagedorn temperature $T_H$. This result was interpreted in the bulk as reflecting the tensionless limit of three distinct strings. The one corresponding with all holographic CFTs (meaning with $a=c$ at large $N$) was identified with the usual 10d Type II string, while a second one was argued to be that proposed in \cite{Gadde:2009dj} to provide the string-theoretic bulk dual of $\mathcal N=2$ superconformal QCD. The nature of the third one, as well as a string-theoretic confirmation of the proposal of \cite{Calderon-Infante:2024oed} following the logic in \cite{Gadde:2009dj}, will be addressed in \cite{to-appear}.

\medskip

In this work, we extend the study of the CFT Distance Conjecture and the bulk interpretation of infinite-distance limits to a broader family of CFTs with \emph{multi-dimensional} conformal manifolds. In particular, we focus on four-dimensional $\mathcal{N}=2$ quiver gauge theories with gauge group
\[
G=SU(N_1)\otimes\cdots\otimes SU(N_p),
\]
and hypermultiplets in fundamental and bifundamental representations. We pursue two main goals: First, we test the interpretation of the \emph{overall} free limit of these quivers as a tensionless-string limit by investigating how their large-$N$ Hagedorn temperature reflects the stringy UV completion of the bulk physics.\footnote{We use the term \emph{stringy UV-completion} as a shorthand for the string theory in which the AdS bulk is embedded. That is, the string theory in which the AdS vacua providing the bulk dual to the gauge theory is most naturally built. For non-holographic CFTs, one cannot think of an Einstein gravity being UV-completed into a string theory, but perhaps of a (partially gauged) higher-spin gravity theory or something more exotic.} For \emph{linear quivers}, we show that their large-$N$ Hagedorn temperatures only depend on the length $p$ and argue that this reflects the stringy UV-completion of their bulk duals as suggested by their Hanany--Witten brane realizations. For \emph{holographic quivers} (that satisfy $a=c$ in the large $N$ and possibly large quiver size limit) we show that the Hagedorn temperature is always that of $\mathcal N=4$ SYM, which neatly fits with the stringy UV-completion of their bulk duals into the usual 10d Type IIB string theory. Second, since these theories possess a (complex) multidimensional conformal manifold, we study the structure of HS towers along different (partial) weak-coupling limits by using convex-hull methods and derive bounds on the corresponding exponential rates $\alpha$. Furthermore, we study the interplay between this parameter $\alpha$ and the large-$N$ Hagedorn temperature for overall weak-coupling limits as proxies for the nature of the HS tower and the UV-completion above its scale. For non-holographic CFTs, we discuss how the relation between these two is subtler than in the case of simple gauge groups.

\paragraph{Outline.}
In Section~\ref{sec:rev}, we review the CFT Distance Conjecture, current bounds and expressions for the exponential rate $\alpha$, and the bulk interpretation of infinite-distance limits as leading to tensionless strings in AdS. In Section~\ref{sec:TH}, we compute the large-$N$ Hagedorn temperature in the overall free limit of four-dimensional $\mathcal{N}=2$ linear and holographic quivers and discuss its relation to the stringy UV-completion in the bulk. Section \ref{sec: alpha} is devoted to the parameter $\alpha$ in the CFT Distance Conjecture. In Section \ref{ss:bounds}, we derive rigorous two-sided bounds on the minimal (overall free) value of the exponential decay rate, $\alpha_{\min}$, in the large $N$ limit and prove the lower bound on $\alpha$ proposed in \cite{Perlmutter:2020buo} for finite $N$. We also analyze weak-coupling directions on the conformal manifold beyond the overall free limit using convex-hull methods in Section \ref{ss:convex-hulls}. Finally, we discuss the relation between $\alpha$ and the Hagedorn scale in Section \ref{ss:alpha-vs-T_H}. We offer some final thoughts and discuss future directions in Section~\ref{sec:conclusion}. An explicit calculation of the Hagedorn temperature for ADE quivers is relegated to Appendix~\ref{App:ADE explicit}, while Appendix \ref{App: HS Density} contains some results regarding the density of the tower of HS states becoming light at weak-coupling infinite-distance limits.

\section{Review: The CFT Distance Conjecture and Tensionless String Limits}\label{sec:rev}

\paragraph{The CFT Distance Conjecture.}
A $d$-dimensional CFT may admit a \emph{conformal manifold}, $\mathcal{M}$, generated by a set of exactly marginal operators, $\mathcal{O}_i$, and parametrized by a set of exactly marginal couplings, $t^i$.\footnote{The existence of an exactly marginal operator in a CFT implies a continuous family of them generated by it. The converse statement is believed to be true and, under the assumption that there exists a conformal interface connecting nearby CFTs within this family, was recently proven in \cite{Komatsu:2025cai}.} There is a natural notion of metric on $\mathcal{M}$ due to Zamolodchikov \cite{zamolodchikov1986irreversibility}, given by the two-point function of the marginal operators
\begin{equation}
    \chi_{ij}(t^k) = |x-y|^{2d} \langle \mathcal{O}_i(x)\,\mathcal{O}_{j}(y)\rangle_{t^k} \, .
\end{equation}
Note that the normalization of this metric is related to that of the marginal operators $\mathcal O_i$.

Motivated by the Distance Conjecture, we are interested in \emph{infinite-distance limits} in $\mathcal{M}$.
All currently known examples of such limits occur in 4d SCFTs and involve (possibly partial)
weak-coupling limits in which a sector of the theory becomes free \cite{Baume:2020dqd,Perlmutter:2020buo}.
In such limits, an infinite tower of higher-spin (HS) currents with spin $J > 2$ becomes conserved, i.e., their anomalous dimensions
\begin{equation}\label{rev: anomalous-dim}
  \gamma_J \;\equiv\; \Delta_J - (J+d-2) \,,
  \qquad \gamma_J \ge 0\,,
\end{equation}
tend to zero, thus leading to an enhancement of the conformal symmetry group to a HS symmetry.\footnote{Having HS symmetry imposes strong constraints on any CFT in $d>2$. In fact, it has been argued that not only any free theory has this symmetry, but also that HS symmetry is always related to a free sector in the theory \cite{Maldacena:2011jn,Stanev:2013qra, Boulanger:2013zza, Alba:2013yda, Alba:2015upa,Hartman:2015lfa, Li:2015itl}.} The inequality $\gamma_J\ge 0$ follows from the unitarity bound. This observation, together with expectations coming from the Distance Conjecture in the bulk, led to the \emph{CFT Distance Conjecture} proposed in \cite{Perlmutter:2020buo}. The conjecture considers a conformal manifold of local, unitary CFTs in $d>2$, and states that
\begin{enumerate}
  \item All CFTs with HS symmetry in the conformal manifold are at infinite distance.
  \item All CFTs at infinite distance in the conformal manifold possess HS symmetry.
  \item The conformal dimensions of the HS currents approach the unitarity bound \emph{exponentially} fast in the geodesic distance computed using the Zamolodchikov metric.
\end{enumerate}
Remarkably, the conjecture does not assume that the CFT has a large central charge, suggesting that the Distance Conjecture in AdS may apply beyond theories that admit an Einstein-gravity description at low energies. Additionally, although all known examples of CFTs with a conformal manifold involve supersymmetry, the conjecture does not assume this ingredient.

The first part of the CFT Distance Conjecture was proved in \cite{Baume:2023msm} by combining the evolution equation for $\gamma_J$, implied by conformal perturbation theory, with the constraints imposed by weakly broken HS symmetry. As suggested by the conjecture, this proof does not assume that the CFT has a large central charge or that it is supersymmetric. The second part remains more challenging to prove, but it is satisfied in all fully controlled examples of CFTs with finite central charge.\footnote{Via its Einstein-gravity $AdS_4$ bulk dual (i.e., in the large central charge expansion), a potential counterexample in $d=3$ has been proposed and investigated in \cite{Bobev:2021yya,Giambrone:2021zvp,Guarino:2021kyp,Cesaro:2021tna,Bobev:2023bxs}. Although there is an infinite tower of exponentially light states in the infinite-distance limit, it does not seem to contain states of arbitrarily high spin. Therefore, this model provides a potential counterexample to the CFT Distance Conjecture, but not to the usual Distance Conjecture in the bulk.} Finally, the third part has been shown to hold in the weak-coupling limits of any 4d SCFT, which necessarily involve free vectors \cite{Perlmutter:2020buo}. In this case, the marginal coupling is identified with the (complexified) gauge coupling, $\tau=\frac{4\pi i}{g^2}+\frac{\theta}{2\pi}$, and the exponential behavior predicted by the Distance Conjecture is closely related to the locally hyperbolic form of the Zamolodchikov metric as $\mathrm{Im}(\tau)\to\infty$ \cite{Baume:2020dqd,Perlmutter:2020buo}. Choosing the overall normalization of this metric such that it measures the moduli-space distance in the bulk with respect to a canonically normalized massless field in AdS and in 5d Planck units, its leading piece reads
\begin{equation}
  ds^2 =  \frac{\dim G_{\rm free}}{8\, c}\,\frac{d\tau\, d\bar{\tau}}{\mathrm{Im}(\tau)^2}\,,
  \label{rev:Zmetric}
\end{equation}
where $\dim G_{\rm free}$ is the dimension of the gauge sector that becomes free as we approach the infinite-distance point and $c$ is the central charge of the \emph{full} CFT. The tower of HS currents becomes conserved as $\gamma_J \sim \mathrm{Im}(\tau)^{-1}$, which in turn translates into an exponentially light tower of HS modes in AdS with a mass scale
\begin{equation}
  m \sim e^{-\alpha \,d(\tau,\tau_0)}\, \quad \text{as} \quad d(\tau,\tau_0)\to \infty \, ,
\end{equation}
where the parameter $\alpha$ is given by \cite{Perlmutter:2020buo}
\begin{equation}\label{rev: alpha}
  \alpha \;=\; \sqrt{\frac{2c}{\dim G_{\rm free}}}\,.
\end{equation}
A basic consequence of \eqref{rev: alpha} is that, among all weak-coupling limits, $\alpha$ is minimized when $\dim G_{\rm free}$ is maximized. For a gauge theory, this happens in the \emph{overall} free limit, in which all gauge couplings are taken to zero at the same rate, namely $g_i \sim g \to 0$. As shown in \cite{Calderon-Infante:2024oed} for any 4d SCFT, this minimum value $\alpha_{\rm min}$ can be expressed purely in terms of the ratio of the $a$ and $c$ central charges as
\begin{equation}
  \alpha_{\min}
  \;=\; \frac{1}{\sqrt{2}}\;\frac{1}{\sqrt{2\,\frac{a}{c}-1}}\, .
  \label{rev:alphamin_ac}
\end{equation}
This shows that all holographic superconformal gauge theories, which necessarily satisfy $a=c$ at large $N$, share the same value of $\alpha$ in their weak-coupling limit.

We will be primarily interested in four-dimensional $\mathcal{N}=2$ superconformal gauge theories, for which all complexified gauge couplings are marginal. Hence, the conformal manifold always contains an overall weak-coupling limit. For this class of theories, the central charges can be written as
\begin{equation}
c=\frac{n_v}{6}+\frac{n_h}{12} \, ,\qquad
a=\frac{5n_v}{24}+\frac{n_h}{24} \, ,\label{rev:acnhnv}
\end{equation}
where $n_h$ and $n_v$ are the number of $\mathcal{N}=2$ hypermultiplets and vector multiplets, respectively. Using these expressions, one can alternatively write \eqref{rev:alphamin_ac} as
\begin{equation}\label{rev:alphanhnv}
    \alpha_{\rm min} = \sqrt{\frac{1}{3} + \frac{1}{6} \frac{n_h}{n_v}} \, .
\end{equation}
This way of writing the minimum value of $\alpha$ makes the lower bound found in \cite{Perlmutter:2020buo} manifest\footnote{Although we have focused on $\mathcal N=2$ superconformal gauge theories for convenience, we stress that the bound derived in \cite{Perlmutter:2020buo} holds for any 4d $\mathcal N=2$ SCFT with a conformal manifold. Moreover, there is a similar lower bound in the 4d $\mathcal N=1$ case.}
\begin{equation}
     \alpha \;>\; \frac{1}{\sqrt{3}}\,,
\end{equation}
where the inequality is strict because a theory with only vector multiplets ($n_h = 0$) does not admit a conformal manifold. Therefore, this bound is not expected to be sharp. Indeed, \cite{Perlmutter:2020buo} proposed the following stronger bound
\begin{equation}
  \alpha \;\ge\; \frac{1}{\sqrt{2}}\,.
  \label{eq:alphabound}
\end{equation}
We would like to note that, for this bound to be universally satisfied, any 4d $\mathcal N=2$ superconformal gauge theory should have $a/c \leq 1$. This is due to \eqref{rev:alphamin_ac} and that these theories always admit an overall-free limit in their conformal manifold.\footnote{There are 4d $\mathcal N=1$ SCFTs with $a/c > 1$ (see e.g. \cite{Intriligator:1994rx,Cho:2024civ,Cho:2025xod}). They are realized as IR fixed points of 4d $\mathcal N=1$ supersymmetric gauge theories and, as shown in \cite{Cho:2024civ,Cho:2025xod}, they can admit a conformal manifold. Nevertheless, the presence of an overall-free limit within this conformal manifold seems unlikely, since in this limit the SCFT should reduce to a far less exotic 4d $\mathcal N=1$ superconformal gauge theory.}

In Section~\ref{ss:bounds}, we will derive strong two-sided bounds on $\alpha_{\min}$ at large $N$ and verify~\eqref{eq:alphabound} (including at finite $N$) for any $\mathcal{N}=2$ superconformal quiver gauge theory. Moreover, in Section~\ref{ss:convex-hulls}, we go beyond the overall free limits by computing the $\vec \alpha$-vectors that encode the exponential decay rate of the HS towers along any weakly coupled direction in the conformal manifold and characterizing their convex hull.

\paragraph{Hagedorn growth and tensionless string limits.}
\label{sec:tensionless_review}
The origin of the tower of light higher-spin (HS) states in the bulk was addressed in \cite{Calderon-Infante:2024oed}, where it was argued that these states arise from the tensionless limit of a string in AdS. This aligns with the Emergent String Conjecture, which states that the light tower emerging at infinite-distance limits is due either to a weakly coupled, tensionless string or to the Kaluza--Klein (KK) modes associated with the decompactification of extra dimensions \cite{Lee:2019xtm}. Since a KK tower does not contain HS states, this motivates the expectation that higher-spin points on conformal manifolds correspond to \emph{tensionless string limits} in the bulk.

A practical diagnostic of stringy behavior is the presence of a Hagedorn-like density of states \cite{Hagedorn:1965st,Hagedorn:1968zz,atick1988hagedorn},
\begin{equation}
  \rho(E)\;\sim\;\exp\!\left(\beta_H E\right)\qquad (E\to\infty)\,,
  \label{eq:Hagedorn_density}
\end{equation}
where $\beta_H=1/T_H$ is the inverse Hagedorn temperature. This exponential growth of the density of states manifests in a divergence of the thermal partition function as the temperature approaches the Hagedorn value, $T \to T_H$ (equivalently, $\beta \to \beta_H$).
This occurs, for example, in the large-$N$ thermal partition function of free gauge theories on $S^3 \times S^1$ \cite{Aharony:2003sx}.\footnote{The large-$N$ limit is required for Hagedorn growth to persist to arbitrarily high energies. At finite $N$, trace relations reduce the number of independent gauge-invariant operators above a certain threshold \cite{Aharony:2003sx}. While this threshold is pushed to parametrically high energies at large $N$, for finite but large $N$ the Hagedorn regime still extends over a wide energy range. From the bulk perspective, this can be viewed as the decoupling of black-hole microstates from the spectrum, allowing the stringy density of states to extend to arbitrarily high energies (see e.g. \cite{Bedroya:2024ubj}).}

Note that the Hagedorn temperature $T_H$ controls the exponential density of states at high energies. It therefore provides a concrete observable that distinguishes different ultraviolet stringy behaviors. In particular, if the weak-coupling limit of a gauge theory is interpreted as a tensionless string limit in the bulk, then different Hagedorn temperatures should diagnose different types of strings. Conversely, one may ask whether two free gauge theories with the same Hagedorn temperature admit the same type of stringy UV completion in the bulk.

This logic was pursued in \cite{Calderon-Infante:2024oed}, where a classification program for infinite-distance limits on the conformal manifold, in terms of the nature of the HS tower in the bulk, was initiated. Motivated by recent results in the study of the Distance Conjecture in string compactifications to flat space, the nature of this tower is expected to be encoded in the parameter $\alpha$. Focusing on 4d superconformal gauge theories with simple gauge group and in the large $N$ limit, three possible values for $\alpha$ were found in \cite{Perlmutter:2020buo}. Given \eqref{rev:alphamin_ac}, this also corresponds to three values for $a/c$. The specific values are
\be \label{eq:types}
(\alpha,\tfrac{a}{c})\in\left\{\left(\tfrac{1}{\sqrt2},1\right),\left(\sqrt{\tfrac{7}{12}},\tfrac{13}{14}\right),\left(\sqrt{\tfrac{2}{3}},\tfrac{7}{8}\right)\right\}.
\ee
Moreover, it was shown in \cite{Calderon-Infante:2024oed} that, at the free point, the large-$N$ Hagedorn temperature of these theories is in one-to-one correspondence with $\alpha$. Consequently, it was argued that these theories fall into three universality classes characterized by these shared properties. Following the logic above, this has been interpreted as signaling three distinct tensionless string limits in AdS. For holographic SCFTs with $a=c$ at large $N$, the corresponding string was identified with the usual 10d Type IIB string. Following the proposal for $\mathcal N=2$ superconformal QCD in \cite{Gadde:2009dj}, the third case in \eqref{eq:types} was identified with a certain 8d Type IIB string theory with a nontrivial worldsheet theory. The nature of the remaining string will be addressed in \cite{to-appear}; there, the proposal of \cite{Calderon-Infante:2024oed} will also be verified by applying the same logic used in \cite{Gadde:2009dj} to the brane constructions of all these theories.

In this work, we push this program beyond simple gauge groups to $\mathcal{N}=2$ conformal quiver theories. In Section~\ref{sec:TH}, we provide evidence for the idea that the large-$N$ Hagedorn temperature of gauge theories diagnoses the type of string theory in the AdS dual by computing the former for several families of free $\mathcal{N}=2$ quiver gauge theories and arguing that it correlates with the expected stringy UV completion in the bulk. For the case of $\mathcal{N}=2$ \emph{linear} quivers, we argue for a certain type of string in the bulk by applying the logic of \cite{Gadde:2009dj,to-appear} to the Hanany--Witten realization of these theories. In agreement with \cite{Calderon-Infante:2024oed}, $\mathcal{N}=2$ \emph{holographic} quivers with $a=c$ at large $N$ (and possibly large quiver size) are shown to share the same Hagedorn temperature (namely, that of $\mathcal{N}=4$ SYM) and are argued to correspond to the usual 10d Type~II string in the bulk. The main difference from \cite{Calderon-Infante:2024oed} is that the one-to-one correspondence between $\alpha$ and $T_H$ is lost for non-holographic theories. We discuss the implications of this in Section~\ref{ss:alpha-vs-T_H}.

\section{Hagedorn Temperature and Stringy UV-completion}\label{sec:TH}

In this section we argue that, for four-dimensional $\mathcal{N}=2$ superconformal quiver gauge theories,
the large-$N$ Hagedorn temperature $T_H$, evaluated in the \emph{overall-free}
(infinite-distance) limit, organizes the theories into universality classes that encode the stringy UV completion of the bulk physics. This quantity captures the exponential growth of the density of high-energy states, which is a common feature of the spectrum of excitations of a string. Our results suggest that this coarse-grained information is enough to determine the string theory in which the bulk is embedded, while forgetting about the detailed geometry in which the string propagates.

We first study \emph{linear quivers} with $SU$ gauge groups in Section~\ref{ss:linear quivers}. We compute the large-$N$
Hagedorn temperature and find a striking universality: $T_H$ depends only on the \emph{length} of the quiver (the
number of gauge nodes) and is completely insensitive to the detailed distribution of ranks and the number of flavors. We then interpret this result in the type IIA Hanany--Witten realization. Following \cite{Gadde:2009dj,to-appear}, we argue that the length of the quiver is tied to the number of NS5-branes and therefore to the type of string theory in which the bulk is embedded. In Section~\ref{ss:holographic-quivers} we turn to \emph{holographic quivers} (again with $SU$ gauge group), by which we mean
large-$N$ $\mathcal{N}=2$ SCFTs with $a \simeq c$. We
show that, for this class of theories, the Hagedorn temperature coincides with that of $\mathcal{N}=4$ SYM,
independently of the detailed quiver data. In the bulk, this naturally reflects the UV-completion given by the fundamental type IIB string.

\subsection{Linear Quivers at Large $N$}\label{ss:linear quivers}
In this subsection, we consider four-dimensional $\mathcal{N}=2$ superconformal quiver gauge theories whose gauge group is a product of $SU(N_i)$ factors arranged in a linear chain. Our first goal is to determine the large-$N$ Hagedorn temperature, $T_H$, of this class of theories from the thermal partition function on $S^3\times S^1$.

\subsubsection{Universality of Hagedorn Temperatures of Linear Quivers}\label{sss:linear}

\begin{figure}[h]
    \centering
    \includegraphics[width=0.45\textwidth]{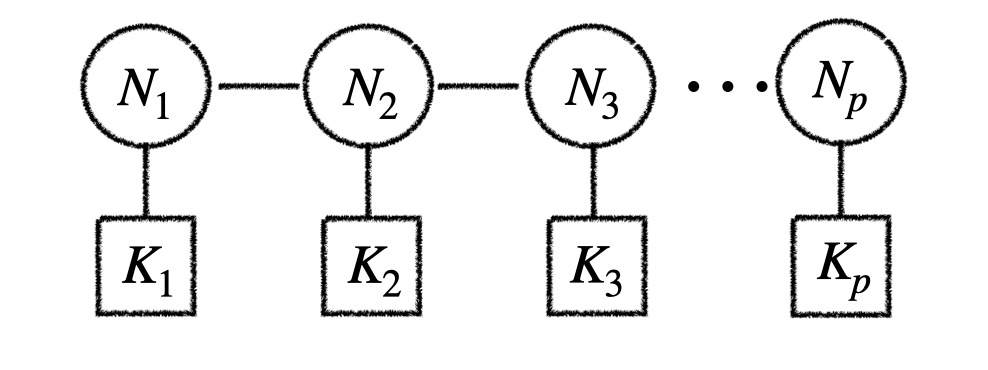}
    \caption{A linear quiver with fundamental flavor multiplets. }
    \label{lquiver}
\end{figure}

Consider a 4d $\mathcal N=2$ linear quiver with the gauge group \( G = SU(N_1) \otimes SU(N_2) \otimes \dots \otimes SU(N_p) \) and hypermultiplets in the bifundamental representations \( N_i \otimes \bar{N}_{i+1} \) for $i=1,\ldots,p-1$. We also include \( K_i \) hypermultiplets in the fundamental representation of the \( i \)-th node. This class of quivers is depicted in Figure \ref{lquiver}. 

\medskip

Following \cite{Sundborg:1999ue,Aharony:2003sx}, we write the thermal partition function of a free $SU(N_1) \otimes \dots \otimes SU(N_p)$ gauge theory on $S^3\times S^1$ as the matrix integral
\begin{equation}
    Z(x) = \int \left(\prod_{i=1}^{p} \mathcal{D}[U_i] \right)\;\text{exp}\left[\sum_{l=1}^{\infty} \sum_{R} \frac{1}{l} z_{R}(x,l) \chi_{R}(u^l)\right] \, .
\end{equation}
The variable $x \equiv e^{-\beta} = e^{-1/T}$ encodes the temperature, while $i = 1,\dots,p\;$ labels the gauge nodes of the quiver and $U_i$ denotes a $SU(N_i)$ matrix with eigenvalues $u_{i,a}$ satisfying $|u_{i,a}|=1$ and $\prod_{a=1}^{N_i} u_{i,a}=1$. The sum over $R$ runs over irreducible representations, and $\chi_R(u)$ denotes the corresponding character which in general can depend on all the eigenvalues, here collectively denoted by $u$. We have also introduced the functions
\begin{equation}
    z_{R}(x,l) = z_{B}^{R}(x^l) + (-1)^{l+1} z_{F}^{R}(x^l),
\end{equation}
that contains the single-particle partition functions for boson and fermions transforming in the irreducible representation $R$. As can be found in \cite{Aharony:2003sx} (see also \cite{Calderon-Infante:2024oed} for a short review), these can be expressed in terms of the number of fields of each type in each representation. Finally, $\mathcal{D}[U_i]$ denotes the \emph{Haar measure} on $SU(N_i)$, which can be written in terms of the eigenvalues of $U_i$ as
\begin{equation} 
\int \mathcal{D}[U_i]=\frac{1}{N_i!}\int \prod_{a=1}^{N_i-1}\frac{d u_{i,a}}{2\pi i u_{i,a}} \left[\Delta(u_i)\Delta(u_i^{-1})\right], 
\end{equation}
where the \emph{Vandermonde determinant} is given by 
\begin{equation} 
\Delta(u_i)=\prod_{1\leq a<b\leq N_i}(u_{i,a}-u_{i,b}) .
\end{equation}
For convenience, we rewrite each integration measure as
\begin{equation} 
    \int \mathcal{D}[U_i]= \frac{1}{N_i!}\int \prod_{a=1}^{N_i-1}\frac{d u_{i,a}}{2\pi i u_{i,a}} \, \text{exp}\left[-\sum_{l=1}^{\infty}\frac{1}{l}\sum_{a\neq b}(u_{i,a})^l(u_{i,b})^{-l}\right]. 
\end{equation}
In this way, the thermal partition function becomes
\begin{equation} 
Z(x)=\prod_{i=1}^{p} \left( \frac{1}{N_i!}\int \prod_{a=1}^{N_i-1} \frac{d u_{i,a}}{2\pi i u_{i,a}} \right) e^{-S_{\text{eff}}}, 
\end{equation}
where the effective action reads
\begin{equation}
S_{\text{eff}} = \sum_{l=1}^{\infty} \frac{1}{l} \Bigg[\sum_{i=1}^{p} \sum_{a\neq b=1}^{N_i}(u_{i,a})^l (u_{i,b})^{-l} - \sum_{R} z_{R}(x,l) \chi_{R}(u^l)\Bigg] \, .
\end{equation}

To this point, our calculations apply to any free $SU(N_1) \otimes \dots \otimes SU(N_p)$ gauge theory. Focusing on $\mathcal N=2$ linear quivers, the effective action can be written as follows
\begin{equation}
\begin{split}
    S_{\text{eff}} = \sum_{l=1}^{\infty} \frac{1}{l} \Bigg[&\sum_{i=1}^{p} \sum_{a\neq b=1}^{N_i}(u_{i,a})^l (u_{i,b})^{-l} - \sum_{i=1}^{p} z_{\text{adj}}(x,l) \chi_{\text{adj}}(u^l_i)\\
    -& \sum_{i=1}^{p-1} z_{\rm biF}(x,l) \left( \chi_{\rm biF}(u^l_{i},u^l_{i+1}) + \chi_{\rm{bi\bar{F}}}(u^l_{i},u^l_{i+1}) \right) \\
    -& \sum_{i=1}^{p} K_i z_{\rm F}(x,l) \left( \chi_{\rm F}(u^l_i) + \chi_{\bar{{\rm F}}}(u^l_i) \right) \Bigg] \, .
\end{split}
\end{equation}
Here we have taken into account that we have the same field content on each adjoint and bifundamental representations (namely, $z_{\text{adj},i}=z_{\text{adj}}$ and $z_{biF,i}=z_{{\rm biF}}$) and $K_i$ copies of a given field content on each fundamental one (that is, $z_{{\rm F},i}=K_i z_{\rm F}$). To keep track of how each representation enters into $S_{\rm eff}$, we are not yet specifying that these field contents actually correspond to a $\mathcal N=2$ vector multiplet for $z_{\text{adj}}$ and a half-hypermultiplet for $z_{\rm biF}$ and $z_{\rm F}$. The characters of these representations are given by
\begin{align}
\chi_{\text{adj}}(u_i) &= \sum_{a,b=1}^{N_i} u_{i,a} (u_{i,b})^{-1} -1,\\
\chi_{\text{biF}}(u_i,u_{i+1}) &= \sum_{a=1}^{N_i} \sum_{b=1}^{N_{i+1}} u_{i,a} (u_{i+1,b})^{-1},\\
\chi_{\text{F}}(u_i) &= \sum_{a=1}^{N_i} u_{i,a},\\
\chi_{\bar{R}}(u) &=\chi_{R}(u^{-1}) \, ,
\end{align}
where the constraint $\prod_{a=1}^{N_i} u_{i,a}=1$ is understood to be implicit in these expressions.

\medskip
 
We are interested in taking the large $N$ limit, in which $N_i \to \infty$. As we will see, our results will not be sensitive to the rates at which the different $N_i$ grow (i.e., to their ratios).\footnote{One could also consider the case in which some $N_i$ stays finite by essentially keeping that node as an spectator in the rest of the computation. For simplicity, we will discuss the case in which all $N_i$ are taken to infinity.} Following \cite{Aharony:2003sx}, in this limit we replace the matrix integral by a path integral over the density of eigenvalues, that is considered to describe a continuum in the large $N$ limit. 

Let us parametrize the eigenvalues as \( u_{i,a}=e^{i \theta_{i,a}} \) with $\sum_{a=1}^{N_i} \theta_{i,a}=0$, and introduce the eigenvalue distributions \( \rho_{i}(\theta_i) \), normalized such that
\begin{equation} \label{eq:rho-normalization}
    \int_{-\pi}^{\pi} \rho_{i}(\theta_i) d \theta_i = N_i \, .
\end{equation}
For any function $f(\theta_{i})$, we have
\begin{equation}
    \sum_{a=1}^{N_i} f(\theta_{i,a}) = \int_{-\pi}^{\pi} d \theta_i\, \rho_{i}(\theta_i) f(\theta_i).
\end{equation}
Using this identity, we rewrite the effective action as\footnote{When rewriting the contribution from the Haar measure, we are ignoring the lack of $a=b$ contribution, which becomes of measure zero in the large $N$ limit in comparison to the $a\neq b$ ones. Alternatively, one may take this into account explicitly, which generates an extra term that is reabsorbed by the definition of the measure for the path integral over $\rho_i$ that will be discussed later on.}
\begin{equation}
\begin{split}
    S_{\text{eff}} = \sum_{l=1}^{\infty} \frac{1}{l} \Bigg[ &\sum_{i=1}^p (1-z_{\text{adj}}(x,l)) \rho^{(l)}_i \bar{\rho}^{(l)}_i - \sum_{i=1}^{p-1} z_{\text{biF}}(x,l) \left( \rho^{(l)}_i \bar{\rho}^{(l)}_{i+1} + \bar{\rho}^{(l)}_i \rho^{(l)}_{i+1} \right) \\
    &- \sum_{i=1}^{p} K_i z_{\rm F}(x,l) \left( \rho^{(l)}_i + \bar{\rho}^{(l)}_i \right) + p \, z_{\text{adj}}(x,l) \Bigg] \, ,
\end{split}
\end{equation}
where
\begin{equation}
    \rho_{i}^{(l)} \equiv \int_{-\pi}^{\pi} d \theta_i\, \rho_{i}(\theta_i) e^{i l \theta_i} 
\end{equation}
are the Fourier modes of the distributions $\rho_i(\theta_i)$ and we have taken into account that $\rho_{i}^{(-l)}=\bar\rho_{i}^{(l)}$.

In the large $N$ limit, we think of the eigenvalues as describing a continuum and hence we treat the densities of eigenvalues $\rho_i(\theta_i)$ as continuous functions. In this continuum limit, we replace the matrix integral by a path integral over $\rho_i(\theta_i)$. More precisely, this path integral is written as an integral over all Fourier modes of $\rho_i$. We obtain
\begin{equation} \label{eq:path-integral}
Z(x)=\prod_{i=1}^{p}\prod_{l=1}^{\infty}\left(\frac{1}{l\pi}\int d^2\rho^{(l)}_{i} \right) e^{-S_{\text{eff}}},
\end{equation}
where the measure for the integrals over $\rho^{(l)}_i$ are fixed by imposing $Z(x)=1$ for $z_{\rm adj}=z_{\rm biF}=z_{\rm F}=0$, that is, by matching the normalization of the Haar measures in the continuum limit. The integral should be performed over the region $|{\rm Re}\,\rho_i^{(l)}| \leq N_i$ and $|{\rm Im}\,\rho_i^{(l)}| \leq N_i$, and then taking the limit $N_i \to \infty$.\footnote{This condition comes from the normalization of $\rho_i$ in \eqref{eq:rho-normalization}.} In most cases, it will suffice to directly think of the integrals as being performed on the entire complex plane.    

With these preparations in place, we are ready to obtain the thermal partition function in the large $N$ limit. In fact, notice that equation \eqref{eq:path-integral} has the form of a multi-dimensional Gaussian integral for each level $l$. However, our main interest is in the Hagedorn temperature $T_H$, namely, the smallest temperature for which $Z(x_H) \to \infty$ (see Section \ref{sec:rev}). Hence, instead of showcasing the solution for $Z(x)$ in the large $N$ limit, let us directly discuss the conditions under which it diverges. For each fixed $l$, the Gaussian integral converges whenever the Hessian of $S_{\rm eff}$ with respect to ${\rm Re}\,\rho_{i}^{(l)}$ and ${\rm Im}\,\rho_{i}^{(l)}$ is a positive-definite matrix. As the temperature grows, this condition fails when the determinant of the Hessian vanishes. In what follows, we will focus on determining the smallest temperature $T_H$ such that this determinant vanishes.

\medskip

Before going on, we need to discuss some subtleties that arise when the numbers of some of the flavors $K_i$ grow in the large $N$ limit. As we will see later, this happens quite generically when imposing the \emph{balancing condition} that ensures that the $\mathcal{N}=2$ quiver gauge theory remains conformal beyond the free point. First, we note that this leads to a divergent $Z(x)$ in the large $N$ limit. From the Gaussian-integral point of view, this comes from the value of the integrand at the saddle point blowing up in this limit. However, as pointed out in \cite{Calderon-Infante:2024oed}, this divergence is of a different nature from the Hagedorn one; the former is not related to an exponentially growing density of states but to the spectrum of the theory not being sparse in the large $N$ limit. Given that we are interested in the stringy exponential growth of the density of states, we factor out the divergence coming from the large number of flavors and focus on the further divergence that happens when $T \to T_H$. As considered in \cite{Gadde:2009dj,Calderon-Infante:2024oed}, it is possible to obtain a sparse spectrum at large $N$ by restricting to the flavor-singlet sector of the theory. We will not explore this direction further in this work.

Another subtlety arises when taking into account that, as explained above, we should first perform the integral in \eqref{eq:path-integral} for $|{\rm Re}\,\rho_i^{(l)}|\leq N_i$ and $|{\rm Im}\,\rho_i^{(l)}|\leq N_i$ and then take $N_i\to\infty$. When $K_i\to\infty$ in this limit, it may happen that the center of the Gaussian is not contained in the integration region. As noticed in \cite{Schnitzer:2004qt}, this can happen above a critical temperature $T_c$. Hence, whenever $T_c<T_H$, the analysis above in terms of a Gaussian integral is not directly applicable. Nevertheless, as argued in \cite{Calderon-Infante:2024oed}, the exponential growth of the density of states that we are interested in is blind to the matter content in the (anti-)fundamental representation. This is actually related to the subtlety discussed in the previous paragraph. For the free theory, the effect of increasing the number of flavors is to enlarge the degeneracy of states at fixed energy due to the increasing flavor symmetry group. This leads to a non-sparse spectrum when $K_i\to\infty$ but does not affect the Hagedorn density of states, once this non-sparseness is properly factored out. For $T<T_c$, the representation of $Z(x)$ as a Gaussian integral makes it easier to factor out the $K_i\to\infty$ divergence, as discussed above, while for $T\geq T_c$ this is more complicated to see at the level of the partition function. In any case, since we know that the number of flavors should not affect the Hagedorn temperature $T_H$, we can safely ignore this subtlety and effectively proceed as if the number of flavors remains finite in the large $N$ limit.

\medskip

Let us proceed with the computation of the large-$N$ Hagedorn temperature. Ignoring the linear and constant terms, we write the effective action as
\begin{equation} \label{eq:Seff-matrix}
S_{\text{eff}} = \sum_{l=1}^{\infty} \bar{\rho}^{(l)} \cdot M^{(l)} \cdot \rho^{(l)} + \cdots \, ,
\end{equation}
where \( M^{(l)} \) are \emph{tridiagonal Toeplitz matrices} given by 
\begin{equation}
M^{(l)} = 
\begin{bmatrix}
1-z_{\text{adj}}(x,l) & -z_{\text{biF}}(x,l) & 0 & \cdots & 0 \\
-z_{\text{biF}}(x,l) & 1-z_{\text{adj}}(x,l) & -z_{\text{biF}}(x,l) & \cdots & 0 \\
0 & -z_{\text{biF}}(x,l) & 1-z_{\text{adj}}(x,l) & \cdots & 0 \\
\vdots & \vdots & \vdots & \ddots & -z_{\text{biF}}(x,l) \\
0 & 0 & 0 & -z_{\text{biF}}(x,l) & 1-z_{\text{adj}}(x,l)
\end{bmatrix}_{p \times p},\label{tridmatrix}
\end{equation}
As explained above, the Hagedorn temperature is determined by the smallest value of $x$ for which the determinant of the Hessian of $S_{\rm eff}$, taken with respect to ${\rm Re}\,\rho_{i}^{(l)}$ and ${\rm Im}\,\rho_{i}^{(l)}$, vanishes. This condition is equivalent to the vanishing of the determinant of any $M^l$. The zeros of the latter are given by \cite{kulkarni1999eigenvalues}
\begin{equation}
    (1 - z_{\text{adj}}(x,l))\;-\;2\,|z_{\text{biF}}(x,l)|\;\cos\!\Bigl(\tfrac{j\,\pi}{p+1}\Bigr)=0 \, , 
    \quad j = 1,\dots,p \, .
\end{equation} 

To find the smallest of these zeros in terms of $0<x<1$, let us specify the matter content encoded in the functions $z_{\text{adj}}(x,l)$ and $z_{\text{biF}}(x,l)$. As mentioned above, these are given by a $\mathcal N=2$ vector and half-hypermultiplet, respectively. 
However, for easier comparison to the results in \cite{Calderon-Infante:2024oed}, let us write down this matter content in terms of $\mathcal N=1$ vector and chiral multiplets. That is
\begin{equation}\label{N=2-single-particles}
\begin{split}
    z_{\rm adj}(x,l) &= z_{v}(x,l) + z_{c}(x,l) \, , \\
    z_{\rm biF}(x,l) &= z_{c}(x,l) \, ,
\end{split}
\end{equation}
where
\begin{equation} \label{N=1-single-particles}
\begin{split}
    z_{v}(x,l) &= z_V (x^l) + (-1)^{l+1} z_F(x^l)  \, , \\
    z_{c}(x,l) &= 2 z_S(x^l) + (-1)^{l+1} z_F(x^l)  \, ,
\end{split}
\end{equation}
and $z_V(x)$, $z_S(x)$ and $z_F(x)$ are the single-particle partition function of a vector, a real scalar and a Weyl fermion, which are given by \cite{Aharony:2003sx}
\begin{equation} \label{single-particles}
  z_S(x) = \frac{x^2 + x}{(1-x)^3} \, , \quad z_F(x) = \frac{4 x^{\frac{3}{2}}}{(1-x)^3} \, , \quad z_V(x) = \frac{6 x^2 - 2 x^3}{(1-x)^3} \,  \, .
\end{equation}
Using the above relations, we get
\begin{equation} \label{Hagedorn-condition}
    z_v(x,l) + \left\{ 1 + 2 \cos\left(\frac{j \pi }{p+1}\right) \right\} z_c(x,l) = 1 \, , \qquad j=1,\ldots,p \, .    
\end{equation}
We note that $z_c$ monotonically increases from zero to infinity in the range $0<x<1$ for any $l$. The same holds for $z_v$ if $l$ is odd, while $z_v$ monotonically decreases from zero to minus infinity if $l$ is even. Thus, the condition above will be met for smaller $x$ the larger the coefficient accompanying $z_c$ is. Taking into account that the cosine appearing in such coefficient is monotonically decreasing in the range $1\leq j\leq p$, the coefficient will be maximized for the smallest $j$, i.e., for $j=1$. All in all, we find that the right condition for the Hagedorn temperature corresponds to\footnote{Even though it will be irrelevant for our purposes, one can also show that the smallest solution to this equation is obtained for $l=1$ using \eqref{N=1-single-particles} and that $z_V(x)$, $z_S(x)$ and $z_F(x)$ are all monotonically increasing from zero to infinity in the range $0<x<1$.}
\begin{equation} \label{Hagedorn-final}
    z_v(x,l) + \left\{ 1 + 2 \cos\left(\frac{\pi }{p+1}\right) \right\} z_c(x,l) = 1 \, .    
\end{equation}
This equation encodes our main conclusion: \emph{the Hagedorn temperature of any free linear quiver with hypermultiplets in the fundamental representations only depends on its length, $p$}. First, it does not depend on the ratios between the ranks of the different nodes, $N_i$, as they are sent to infinity in the large-$N$ limit. Second, it does not depend on the number of hypermultiplets in the fundamental representations, $K_i$, as already discussed above (see also \cite{Calderon-Infante:2024oed}). Notice that this conclusion does not rely on the gauge theory being conformal (or even UV-complete) beyond the free point. In the next section, we argue that this conclusion reflects that the stringy UV-completion in the bulk---i.e., the string theory in which it is embedded---only depends on the length of the quiver as well.

Before finalizing this section, let us note that \eqref{Hagedorn-final} is similar to the expression that determines the Hagedorn temperatures of superconformal gauge theories with simple gauge groups found in \cite{Calderon-Infante:2024oed}. In fact, theories of type 1, 2, and 3 in the classification of \cite{Calderon-Infante:2024oed} yield the condition above for $p\to\infty$, $p=2$, and $p=1$, respectively. The latter coincidence can be understood as coming from the $p=1$ linear quiver theory reducing to $\mathcal N=2$ $SU(N)$ SQCD with $K$ flavors. For $K=2N$ this coincides with one of the theories of type 3 in \cite{Calderon-Infante:2024oed}. The reason behind the coincidence for the other two cases will be explained in the next section.

\subsubsection{Stringy UV-completion from Hanany-Witten Brane Models} \label{sss:Hanany-Witten}

\begin{figure}[h]
    \centering
    \includegraphics[width=10 cm]{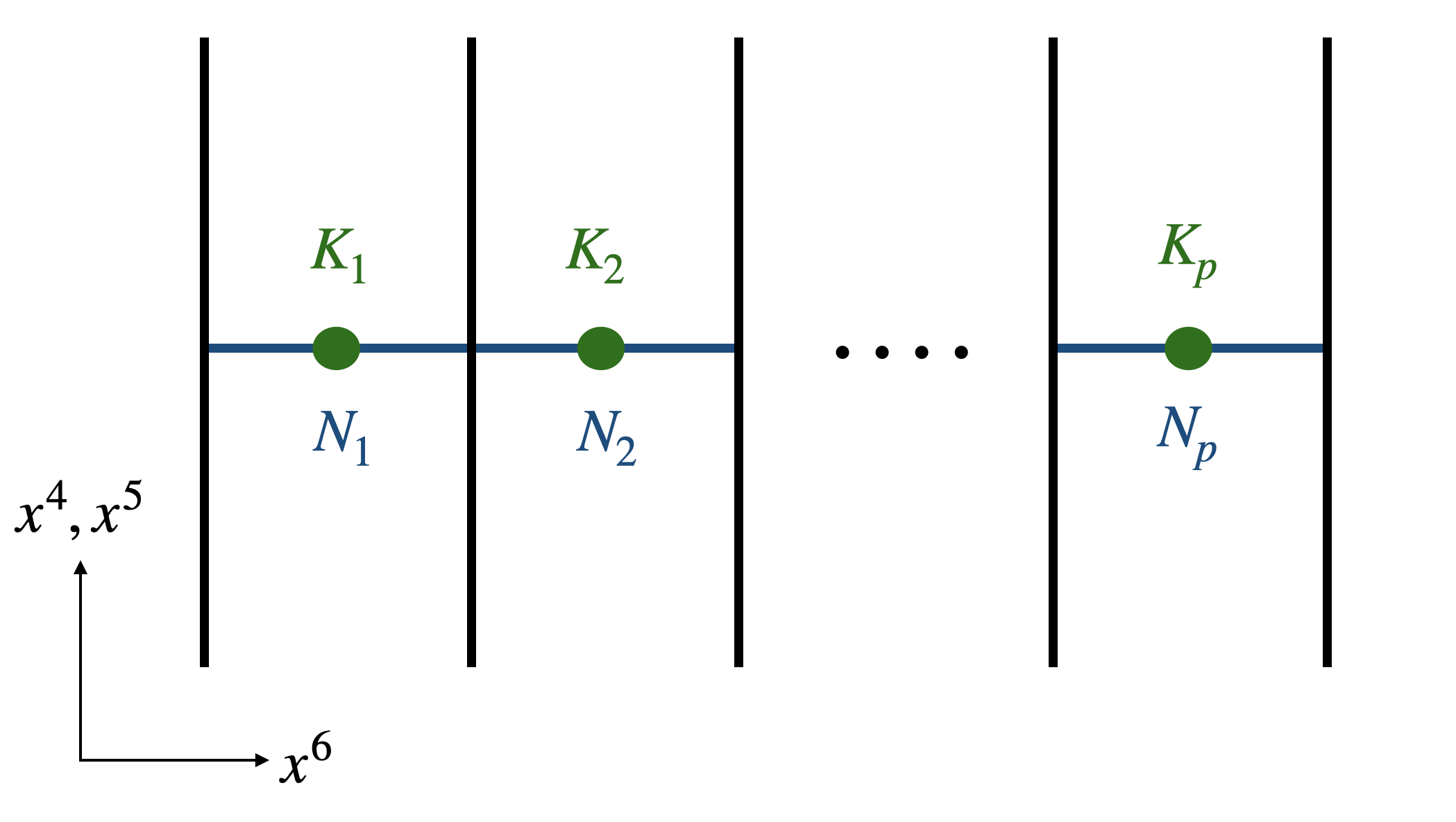}%
\caption{The Hanany--Witten setup. The black lines denote NS5-branes, the blue lines represent a stack of $N_i$ D4-branes, and the green dots represent a stack of $K_i$ D6-branes. }
  \label{fig:hw}
\end{figure}

The $\mathcal N=2$ linear quivers considered in the previous section can be realized via a Hanany--Witten brane configuration in type IIA string theory \cite{Hanany:1996ie}. Concretely, we consider $p+1$ parallel NS5-branes extended along $(x^0,x^1,x^2,x^3,x^4,x^5)$, with $N_i$ D4-branes stretched along $(x^0,x^1,x^2,x^3,x^6)$ between the $i$-th and $(i+1)$-th NS5-branes, and $K_i$ D6-branes extended along $(x^0,x^1,x^2,x^3,x^7,x^8,x^9)$ and localized between the $i$-th and $(i+1)$-th NS5-branes (see Figure~\ref{fig:hw}). This configuration preserves eight supercharges. Strings stretched within the $i$-th stack of D4-branes give rise, in the low-energy limit, to an $SU(N_i)$ vector multiplet. Strings stretched between D4-branes in adjacent stacks give hypermultiplets transforming in the $N_{i-1}$--$\bar{N}_i$ bifundamental representation. In addition, strings stretched between the D4- and D6-branes in the $i$-th interval give rise to $K_i$ hypermultiplets in the fundamental representation of $SU(N_i)$.

The four-dimensional gauge theories under consideration are recovered in the low-energy limit of this configuration. More precisely, we need to take the following limit (see also \cite{to-appear})
\begin{equation} \label{eq:decoupling-limit}
\begin{split}
	&l_s \to 0 \, , \quad |x_{i+1}^6 - x_i^6| \to 0\, , \quad g_s \to 0 \, , \\
	&\text{with} \quad \frac{1}{g_{i}^2} \sim \frac{|x_{i+1}^6 - x_i^6|}{g_s\,l_s} \quad \text{fixed} \, .
\end{split}
\end{equation}
Taking the string length $l_s$ to zero is required to decouple all the stringy modes, while sending the distance between each pair of NS5-branes to zero implements the dimensional reduction of the 5d worldvolume theory living on each stack of D4-branes on the intervals defined by the NS5-branes. To be able to treat these intervals as fixed during the dimensional reduction, the NS5-branes need to be infinitely heavy with respect to the D4-branes. This is achieved by sending the string coupling to zero. Finally, the last line in \eqref{eq:decoupling-limit} ensures that the 4d gauge couplings stay fixed in the limit.

Following the standard logic for obtaining AdS/CFT pairs from string theory \cite{Maldacena:1997re}, one expects the closed-string description of this configuration---in which the branes are replaced by their backreaction on the closed-string degrees of freedom---to admit a (quasi-)AdS\footnote{An exactly AdS near-horizon limit is expected only when the dual gauge theory is conformal.} near-horizon limit, yielding the bulk dual of the corresponding gauge theory. Constructing such a closed-string background and extracting the associated AdS vacua, however, is not straightforward. Rather than pursuing this route, we will follow the approach of \cite{Gadde:2009dj} and argue that the string theory in which the bulk dual is embedded (i.e., UV-completed) is determined by the quiver length $p$.

The authors of \cite{Gadde:2009dj} considered $\mathcal N=2$ superconformal QCD, which corresponds to the linear quiver with $p=1$ and $K=2N$. The Hanany--Witten realization of this theory then involves $N$ D4-branes stretching between two NS5-branes and $2N$ D6-branes placed between them. Alternatively, one can replace the $2N$ D6-branes by two stacks of $N$ semi-infinite D4-branes attached to each NS5-brane and stretching to infinity.\footnote{Starting from the former configuration with $2N$ D6-branes between the NS5-branes, these two realizations are related by moving two stacks of $N$ D6-branes to infinity across each of the NS5-branes. The two stacks of $N$ semi-infinite D4-branes are then generated by the Hanany--Witten effect \cite{Hanany:1996ie}.}  The authors of \cite{Gadde:2009dj} observed that the limit in \eqref{eq:decoupling-limit} coincides with the double-scaling limit of little string theory discussed in \cite{Giveon:1999px,Giveon:1999tq}. In this limit, after performing a T-duality, the near-horizon geometry sourced by the two NS5-branes yields the non-trivial type IIB background described by the worldsheet theory
\begin{equation}\label{eq:non-critical}
    \mathbb R^{1,5} \times SL(2,\mathbb{R})_2/U(1)/\mathbb Z_2 \, .
\end{equation}
The stacks of D4-branes in the Hanany--Witten setup then become $N$ D3-branes and $2N$ D5-branes in the worldsheet theory \eqref{eq:non-critical}. The proposal of \cite{Gadde:2009dj} is that the backreaction of these branes produces a geometry whose near-horizon limit furnishes the AdS dual of $\mathcal N=2$ superconformal QCD.\footnote{See \cite{Dei:2024frl} for recent progress on constructing this background from a supergravity truncation of the string theory in \eqref{eq:non-critical}.} In this way, the bulk dual of this gauge theory is most naturally embedded in the string theory with worldsheet description \eqref{eq:non-critical}. The main point we wish to emphasize is that this background string theory arises solely from the near-horizon geometry of the NS5-branes, while the D-branes appear as additional defects whose backreaction is expected to select the particular (quasi-)AdS vacuum dual to the four-dimensional gauge theory.

By repeating the same argument for the more general Hanany--Witten setup introduced above, we are led to propose that the bulk duals of $\mathcal N=2$ linear quivers are naturally embedded in the string theory obtained from the backreaction of the $p+1$ parallel NS5-branes in the limit \eqref{eq:decoupling-limit}. We will not attempt to describe this background here. For our purposes, the crucial point is that only the number of NS5-branes is relevant for determining the type of stringy UV completion of the bulk. This is closely related to the fact that their tension is parametrically larger than that of the D$p$-branes as $g_s \to 0$, as required above. Consequently, in taking the near-horizon limit of the Hanany--Witten setup in the regime \eqref{eq:decoupling-limit}, one expects to feel the backreaction of the NS5-branes well before that of the D-branes. In this sense, the putative (quasi-)AdS near-horizon region \emph{lives} inside the near-horizon geometry generated by the NS5-branes alone. 

As anticipated, this conclusion fits neatly with the result of the previous section: the large-$N$ Hagedorn temperature of $\mathcal N=2$ linear quivers depends only on $p$. From the viewpoint of the Hanany--Witten brane setup, $p$ is determined by the number of NS5-branes. Thus, the Hagedorn temperature provides a useful diagnostic of the type of string theory in which the bulk is embedded.

\medskip

Finally, let us return to the special values $p=2$ and $p \to \infty$ mentioned at the end of the previous section. We now see that the $p=2$ case, for which the Hagedorn temperature coincides with that of type~2 theories in \cite{Calderon-Infante:2024oed}, is related to the near-horizon limit of three parallel NS5-branes in the limit \eqref{eq:decoupling-limit}. Here we are anticipating results from \cite{to-appear}, where a Hanany--Witten realization involving precisely three NS5-branes is found for all theories of type~2 in \cite{Calderon-Infante:2024oed}. For $p \to \infty$, the Hagedorn temperature converges to that of $\mathcal{N}=4$ SYM (and all type~1 theories in \cite{Calderon-Infante:2024oed}). This leads to the expectation that, in this limit, the bulk theory can be embedded into the usual 10d type~II string, which nicely fits with the results of \cite{Aharony:2012tz}. As observed in \cite{Nunez:2023loo}, these theories have $a=c$ at large $N$ in this limit, which also suggests that they are holographic (in the sense of having weakly coupled Einstein duals). In the next section, we explore the idea that gauge theories with $a \simeq c$ have Hagedorn temperatures close to that of $\mathcal{N}=4$ SYM and admit a 10d Type~II bulk dual beyond linear quivers.

\subsection{The Hagedorn Temperature of Holographic Quivers} \label{ss:holographic-quivers}

In this section, we restrict our attention to holographic quivers, by which we mean $\mathcal{N}=2$ superconformal quiver gauge theories with $a=c$ at large $N$ (and possibly also in the large-quiver limit).\footnote{Recall that having $a=c$ at large $N$ is a necessary condition for a 4d CFT to be holographic, i.e., to have an Einstein gravity bulk dual \cite{Henningson:1998gx,Nojiri:1998dh,Nojiri:1999mh}.} Our goal is to show that their Hagedorn temperature is always that of $\mathcal N=4$ SYM in this limit. This greatly generalizes a similar result found in \cite{Calderon-Infante:2024oed} for any holographic superconformal gauge theory with simple gauge group.

First, let us classify theories with $a \simeq c$. Equivalently, we can focus on $n_h \simeq n_v$, where $n_h$ and $n_v$ denote the numbers of hypermultiplets and vector multiplets, respectively. To do this, let us consider a general $\mathcal N=2$ quiver with $\prod_{i=1}^p SU(N_i)$ gauge group, $A_{ij}$ hypers in the bifundamental representation of $SU(N_i)\times SU(N_j)$,\footnote{Thanks to $\mathcal N=2$ supersymmetry, this hyper is in fact two chiral multiplets in the bifundamental and antifundamental. For this reason, the matrix $A_{ij}$ is in fact symmetric. As the notation suggests, it is nothing but the adjacency matrix of the quiver.} and $K_i$ flavor hypermultiplets in the fundamental representation of $SU(N_i)$.  The number of vector multiplets and hypermultiplets read
\begin{equation} \label{eq:nv&nh}
\begin{split}
    n_v =& \sum_{i=1}^{p} (N_i^2 -1) \simeq \sum_{i=1}^{p} N_i^2 = |\vec N|^2 \, , \\
    n_h =& \sum_{i<j} N_i A_{ij} N_j + \sum_{i=1}^{p} N_i K_i = \frac{1}{2} \vec N \cdot A \cdot \vec N + \vec N \cdot \vec K \, .
\end{split}
\end{equation}
In the first line we have already taken $N_i\gg 1$. To shorten notation, we have also introduced the vectors $\vec N$ and $\vec K$ and written everything in terms of dot products. Let us also recall the \emph{balancing condition}, which in terms of the \emph{Cartan matrix of the quiver} M is given by
\begin{equation} \label{eq:cartan-matrix}
    M \cdot \vec N = \vec K \, , \qquad M \equiv 2 \, \mathbf{I}_p - A \, .
\end{equation}
As can be found e.g. in \cite{Nunez:2023loo}, this condition guarantees that the one-loop beta functions of all the gauge couplings vanish, so that the quiver represents a SCFT at the interacting level. Using the equations for $n_v$ and $n_h$ above, as well as the balancing condition, one finds the identity
\begin{equation} \label{eq:nh/nv-quivers}
    \frac{n_h}{n_v} = 1 + \frac{1}{2} \frac{\vec N \cdot M \cdot \vec N}{|\vec N|^2} = 1 + \frac{1}{2} \frac{\vec N \cdot \vec K}{|\vec N|^2} \, .
\end{equation}
From this equation, we clearly see that having $n_h \simeq n_v$ requires $\vec N \cdot \vec K \ll |\vec N|^2$. This hierarchy can be achieved in two different ways, depending on whether we keep $p$ fixed or allow for a very large quiver.

\paragraph{Holographic quivers at fixed $p$:} If we keep $p$ fixed, the condition above must be satisfied as $N_i \to \infty$. Given that $N_i, K_i \geq 0$, this requires $K_i \ll N_i \quad \forall i$. In other words, for each gauge node, the number of flavors must be negligible with respect to the number of colors. Plugging this into the balancing condition, we see that, at leading order as $N_i \to \infty$, we in fact have $M \cdot \vec N = 0$. This is the condition that leads to the affine ADE quivers \cite{Katz:1997eq}. Even though the holographic requirement leads to this condition only at leading order in the $N_i \to \infty$ limit, once $M$ is fixed to be the Cartan matrix of an affine ADE quiver, the balancing condition \eqref{eq:cartan-matrix} completely fixes $\vec K=0$ and $\vec N$ to be proportional to the vector of affine Dynkin marks.\footnote{We would like to thank Florent Baume for pointing this out to us.}

We conclude that all holographic theories with $p$ fixed correspond to the affine ADE quiver theories. As we show below, the large-$N$ Hagedorn temperature of all these theories coincide with that of $\mathcal N=4$ SYM. This dovetails neatly with the proposal that $T_H$ determines the type of string theory in the bulk: as is well known for $\mathcal N=4$ SYM \cite{Maldacena:1997re}, all of these theories admit ten-dimensional Type IIB holographic duals \cite{Kachru:1998ys,Lawrence:1998ja}.\footnote{Let us stress that, even though these theories are known to admit a 10d Type IIB description, finding a suitable worldsheet theory in the tensionless string limit is far from obvious. In this regime, the background is not weakly-curved in string units and hence the usual $\alpha^\prime$-expansion in the worldsheet breaks down in this regime. See however \cite{Gaberdiel:2021qbb,Gaberdiel:2021jrv,Gaberdiel:2022iot} for recent proposals for the worldsheet duals of free $\mathcal N=4$ SYM and free $\mathcal N=2$ A-type circular quiver} The equality of Hagedorn temperatures with that of $\mathcal N=4$ SYM was already established for the A-type quivers in \cite{Larsen:2007bm}. Using the same method as in Section \ref{sss:linear}, we extend the result to the remaining D-type and E-type quivers in Appendix~\ref{App:ADE explicit}. At the end of this section, we give a more elegant derivation that applies directly to any holographic quiver, including those requiring $p\rightarrow \infty$ discussed next.

\paragraph{Holographic quivers for $p\to\infty$:} Allowing for the quiver to be very large opens up a new way of achieving $\vec N \cdot \vec K \ll |\vec N|^2$. The idea is simple: some of the gauge nodes are allowed to have an order $N$ number of flavors, as long as the number of these nodes is subleading as $p\to \infty$. That is, $K_i \ll N_i$ is satisfied for all nodes with a few exceptions. Examples of this are the large linear quivers studied in Section \ref{ss:linear quivers}. 

To see how this mechanism yields a holographic quiver with $a=c$ at leading order in the large $N$ and large $p$ limit, let us partition the nodes into two subsets, $i \in B$ and $i \notin B$, by defining
\begin{equation}
  B = \left\{ i : \frac{K_i}{N_i} \to 0 \text{ as } N_i \to \infty \right\} \, .
\end{equation}
Then, we can write
\begin{equation}
  \frac{\vec N \cdot \vec K}{| \vec N |^2}
  = \frac{\sum_i N_i K_i}{\sum_i N_i^2}
  = \sum_{i\in B}\frac{K_i}{N_i} \frac{N_i^2}{\sum_j N_j^2}
    + \sum_{i\notin B}\frac{K_i}{N_i} \frac{N_i^2}{\sum_j N_j^2}.
\end{equation}
The first term vanishes in the $N_i \to \infty$ limit by the definition of $B$. For the second term, we use that the balancing condition \eqref{eq:cartan-matrix} does not allow $K_i \gg N_i$.\footnote{Intuitively, taking $K_i \gg N_i$ would make the contribution from the hypermultiplets in the fundamental representations overwhelm that of the vector multiplets in the one-loop beta function. This cannot be compensated by hypermultiplets in bifundamental representations, since they contribute with the same sign as those in the fundamental.} In particular, there exists a constant $c$ such that $K_i/N_i \le c$ for all $i$ as $N_i \to \infty$. Using this bound, we obtain
\begin{equation}
  \frac{\vec N \cdot \vec K}{ |\vec N |^2}
  \leq 
  c\,\frac{\sum_{i\notin B} N_i^2}{\sum_i N_i^2} \, .
\end{equation}
Finally, we note that the balancing condition forces all ranks to be of the same order as $N_i \to \infty$.\footnote{Otherwise, there would exist a gauge node for which the contribution of bifundamental hypermultiplets to its one-loop beta function overwhelms that of the vector multiplet. This cannot be compensated by hypermultiplets in the fundamental representation, since their contribution has the same sign as that of the bifundamentals.} Taking this into account, we see that the right-hand side of the expression above tends to zero as $p \to \infty$ provided that $|\bar{B}| \ll p$ in this limit, i.e.\ provided that only an $\mathcal{O}(1)$ number of nodes do not belong to $B$. Consequently, we have $\vec N \cdot \vec K \ll \lvert \vec N \rvert^2$ in the large $N$ and large quiver limit, which is the condition for holographic quivers discussed above. 

\paragraph{Hagedorn temperature of holographic quivers:} After having characterized all possible $\mathcal{N}=2$ superconformal holographic quivers---namely, those with $a=c$ at large $N$ (and possibly also at large $p$)---we now show that their Hagedorn temperature always coincides with that of $\mathcal{N}=4$ SYM. For later convenience, we recall that the latter is determined by the condition $z_v(x,l) + 3\,z_c(x,l) = 1$ (see \cite{Calderon-Infante:2024oed}), where, as discussed in Section \ref{sss:linear}, $z_v$ and $z_c$ encode the matter content of an $\mathcal N=1$ vector and chiral multiplet, respectively.

Consider a general $\mathcal{N}=2$ quiver gauge theory with $SU$ gauge groups and hypermultiplets in the bifundamental and fundamental representations. The large-$N$ thermal partition function can be computed as in \eqref{eq:path-integral}. Up to terms that are linear or constant (and hence irrelevant for the Hagedorn divergence), the effective action takes the form \eqref{eq:Seff-matrix}, with
\begin{equation}\label{eq:M-general-quiver}
	M^{(l)} = \bigl(1-z_{\rm adj}(x,l)\bigr)\,\mathbf{I}_p \;-\; z_{\rm biF}(x,l)\,A \, ,
\end{equation}
where $A$ is the adjacency matrix of the quiver. As argued in Section~\ref{sss:linear}, the Hagedorn temperature corresponds to the smallest value of $x$ for which $\det M^{(l)}=0$ for some $l$. Equivalently, it occurs when the smallest eigenvalue of $M^{(l)}$ reaches zero. Using \eqref{eq:M-general-quiver}, this yields
\begin{equation}
	1 - z_{\rm adj}(x,l) - \lambda_{\max}(A)\, z_{\rm biF}(x,l) = 0 \, ,
\end{equation}
where $\lambda_{\max}(A)$ is the largest eigenvalue of $A$. Substituting \eqref{N=2-single-particles} into the previous equation, we obtain
\begin{equation} \label{eq:Hagedorn-condition-general}
	z_v(x,l) + \bigl(1+\lambda_{\max}(A)\bigr)\,z_c(x,l) = 1 \, .
\end{equation}
Interestingly, all dependence on the particular $\mathcal N=2$ quiver is thus encoded in $\lambda_{\max}(A)$.\footnote{For linear quivers one has $\lambda_{\max}(A)=2\cos\!\left(\frac{\pi}{p+1}\right)$, hence recovering \eqref{Hagedorn-final}.} Notice that the condition for the Hagedorn temperature of $\mathcal N=4$ SYM recalled above is recovered when $\lambda_{\max}(A)=2$. 

Since $A$ is a real symmetric matrix with non-negative entries and $\vec N$ is a strictly positive vector, the Collatz--Wielandt bounds and the Rayleigh-quotient bound imply \cite{horn2012matrix} 
\begin{align}
	\min_i \frac{(A\cdot \vec N)_i}{N_i}
	\;\le\;& \lambda_{\max}(A) \;\le\;
	\max_i \frac{(A\cdot \vec N)_i}{N_i}\, ,
	\\[2pt]
	\frac{\vec N \cdot A \cdot \vec N}{|\vec N|^2}
	\;\le\;& \lambda_{\max}(A)\, .
\end{align}
For any $\mathcal N=2$ \emph{superconformal} quiver, the balancing condition implies $A\cdot \vec N = 2\vec N - \vec K$. Substituting this into the previous inequalities yields
\begin{align}
	2-\max_i\frac{K_i}{N_i}
	\;\le\;& \lambda_{\max}(A) \;\le\;
	2-\min_i\frac{K_i}{N_i}\,,\label{CW}
	\\
	2-\frac{\vec N\cdot \vec K}{|\vec N|^2}
	\;\le\;& \lambda_{\max}(A) \, .\label{Rei}
\end{align}
We note that the upper bound on $\lambda_{\max}(A)$ implies that the Hagedorn temperature of any $\mathcal N=2$ superconformal quivers is larger or equal to that of $\mathcal N=4$ SYM. As discussed above, for \emph{holographic} quivers we furthermore have
\begin{equation}
	\min_i\frac{K_i}{N_i}\to 0
	\qquad\text{and}\qquad
	\frac{\vec N\cdot \vec K}{|\vec N|^2}\to 0 \, ,
\end{equation}
in the large $N$ limit (and, if applicable, also in the large $p$ limit). Hence, the upper bound \eqref{CW} together with the lower bound \eqref{Rei}, squeeze the allowed range for $\lambda_{\max}(A)$, yielding
\begin{equation}
	\lambda_{\max}(A)\;\rightarrow\;2 \, .
\end{equation}
As advertised, this shows that the Hagedorn temperature of any $\mathcal N=2$ superconformal holographic quiver coincides with that of \(\mathcal N = 4\) SYM. Following the logic that $T_H$ determines the stringy UV-completion in the bulk, this suggests that all these holographic SCFTs admit a dual description in 10d Type IIB string theory.

\section{Leading Tensionless String and the CFT Distance Conjecture}\label{sec: alpha}

In this section, we turn our attention to the CFT Distance Conjecture in the (partial) weak-coupling limits of $\mathcal N=2$ quiver SCFTs with hypermultiplets in the bifundamental and fundamental representations. Our goal is twofold. First, we explore the possible values of the parameter $\alpha$ in the CFT Distance Conjecture. In the overall free limit—namely, when all gauge couplings go to zero at the same rate—we derive two-sided bounds on this parameter. To go beyond this particular direction in the conformal manifold, we compute the so-called $\vec\alpha$-vectors and discuss the geometry of the convex hull that they generate. Second, we connect with our discussion of Hagedorn temperatures and stringy UV completion. By explicit counterexamples, we find that the one-to-one correspondence between $\alpha$ and $T_H$ observed in \cite{Calderon-Infante:2024oed} does not hold beyond simple gauge groups. We then explain why this does not necessarily contradict the common lore that $\alpha$ encodes the nature of the leading tower. Throughout this section, we highlight similarities and differences between our findings in AdS and in flat space.

\subsection{Bounds on the Distance Parameter for $\mathcal{N}=2$ Quivers} \label{ss:bounds}

The goal of this section is to derive bounds on the parameter $\alpha$ that appears in the CFT Distance Conjecture. In particular, we focus on the overall free limit of $\mathcal N =2$ quivers with hypermultiplets in the fundamental and bifundamental representations. This overall free limit---for which all the gauge couplings go to zero at the same rate, namely $g_i \sim g \to 0$---is known to yield the smallest possible value of $\alpha$ in the conformal manifold (see Section \ref{sec:rev}). For the class of quivers at hand, we will derive two-sided bounds on $\alpha_{\rm min}$ in the large $N$ limit.

Our strategy is to bound the ratio of the number of hypermultiplets to vector multiplets, $r \equiv n_h/n_v$. The central charges $a$ and $c$ are given in \eqref{rev:acnhnv}. It then follows that
$
\frac{a}{c}=\frac{5+r}{4+2r}\,.
$
Since this ratio is strictly decreasing for $r\ge 0$, a lower bound on $r$ implies an upper bound on $a/c$, and conversely. Using \eqref{rev:alphanhnv}, we will translate these into bounds on $\alpha_{\rm min}$. Finally, we discuss how our results generalize to finite $N$, yielding an absolute lower bound on $\alpha$.

\medskip

Let us first focus on the lower bound on $\alpha_{\rm min}$, which is equivalent to an upper and a lower bound on $a/c$ and $n_v/n_h$, respectively. This bound follows from equation \eqref{eq:nh/nv-quivers}, which we recall here for convenience
\begin{equation} \label{eq:nh/nv-quivers-2}
    \frac{n_h}{n_v} = 1 + \frac{1}{2} \frac{\vec N \cdot M \cdot \vec N}{|\vec N|^2} = 1 + \frac{1}{2} \frac{\vec N \cdot \vec K}{|\vec N|^2} \, .
\end{equation}
Since both $\vec N$ and $\vec K$ have non-negative entries, we have $\vec N \cdot \vec K \ge 0$. Thus, we obtain the bounds
\begin{equation} \label{eq:lower-bounds}
    \frac{n_h}{n_v} \geq 1 \, , \qquad  \frac{a}{c} \leq 1 \, , \qquad \alpha_{\rm min} \geq \frac{1}{\sqrt{2}} \, ,
\end{equation}
in the large $N$ limit. Notice that this bound is saturated for $\vec K = 0$, i.e.\ in the large-$N$ limit of the ADE quivers discussed in Section~\ref{ss:holographic-quivers}.

Similarly, we can bound \(n_v/n_h\) from above by rewriting \eqref{eq:nh/nv-quivers-2} and substituting the definition of the quiver's Cartan matrix, \(M \equiv 2\,\mathbf{I}_p - A\). This yields
\begin{equation} 
    \frac{n_h}{n_v} = 2 - \frac{1}{2} \frac{\vec N \cdot A \cdot \vec N}{|\vec N|^2} \, .
\end{equation}
Recall that \(A\) is the adjacency matrix, so all of its entries are non-negative. Therefore, the second term in the equation above is also non-negative, and we obtain the bounds
\begin{equation} \label{eq:upper-bounds}
    \frac{n_h}{n_v} \leq 2 \, , \qquad  \frac{a}{c} \geq \frac{7}{8} \, , \qquad \alpha_{\rm min} \leq \sqrt{\frac{2}{3}} \, ,
\end{equation}
in the large $N$ limit. In this case, the bound is saturated when $A=0$. This corresponds to a single-node quiver, i.e., $\mathcal{N}=2$ SCQCD.

\medskip

The two-sided bounds we have just derived apply to the leading large-$N$ behavior of $n_v/n_h$, $a/c$, and $\alpha_{\rm min}$. To determine whether these bounds extend to finite $N$, we recall that the only large-$N$ approximation occurs in deriving \eqref{eq:nh/nv-quivers}. This approximation overestimates $n_v$, and therefore \emph{underestimates} $n_h/n_v$. Consequently, the bounds in \eqref{eq:upper-bounds} need not hold away from the large-$N$ limit. By contrast, the bounds in \eqref{eq:lower-bounds} remain valid at finite $N$. Moreover, since we have a lower bound on the smallest value of $\alpha$ (namely $\alpha_{\min}$) along the conformal manifold of a given theory, it can be promoted to an absolute lower bound on $\alpha$
\begin{equation} \label{eq:lower-bound-finiteN}
    \alpha \geq \frac{1}{\sqrt{2}} \quad \text{(finite $N$)} \, ,
\end{equation}
which coincides with the bound proposed in \cite{Perlmutter:2020buo}. Our results verify this bound for any $\mathcal{N}=2$ superconformal quiver gauge theory, even at finite $N$!

\subsection{Convex Hulls for Weakly-coupled Gauge Theories} \label{ss:convex-hulls}

In this section, we aim to characterize the value of $\alpha$ beyond the \emph{overall} weak-coupling limit considered in the previous section. As is customary in studies of the Distance Conjecture, we do so by computing the so-called scalar charge-to-mass ratios, or $\vec{\alpha}$-vectors, and by characterizing their convex hull. In particular, we compute these vectors for the various towers of HS states in the \emph{completely weakly coupled} regime, namely for $g_i \ll 1$. This can be viewed as a particular \emph{duality frame} on the conformal manifold. As in the moduli space of string compactifications (see, e.g., \cite{Calderon-Infante:2020dhm,Etheredge:2022opl,Etheredge:2024tok,Grieco:2025bjy}), we will see that this yields what can be identified as a \emph{frame simplex}. 

\medskip

Following \cite{Calderon-Infante:2020dhm}, we define the $\vec{\alpha}$-vectors as
\begin{equation}
	\vec{\alpha}_{\rm t} \equiv - \vec{\nabla} \log\!\left(\frac{M_{\rm t}}{M_{\rm Pl}}\right)\, ,
\end{equation}
where $M_{\rm t}$ is the mass scale of the tower, and the derivatives are taken with respect to an orthonormal basis in moduli space. Given a trajectory approaching an infinite-distance point with unit tangent vector $\hat{T}$, we can compute the exponential rate at which the tower becomes light as
\begin{equation}
	\alpha_{\rm t} = \hat{T} \cdot \vec{\alpha}_{\rm t} \, .
\end{equation}
Hence, the $\vec{\alpha}$-vectors contain the relevant information for the Distance Conjecture. More precisely, the convex hull of these vectors has been shown to encode, geometrically, the features of different asymptotic regimes.

\medskip

To compute these vectors from the CFT, we first need the Zamolodchikov metric. For 4d $\mathcal{N}=2$ theories, it can be computed from the four-sphere partition function via \cite{Gerchkovitz:2014gta,Gomis:2014woa,Baggio:2014ioa}
\begin{equation}
	\chi_{i \bar \jmath} \sim \partial_i \partial_{\bar \jmath} \log Z_{S^4} \, ,
\end{equation}
where the $\mathcal{O}(1)$ factor depends on the normalization chosen for the Zamolodchikov metric.
 For a gauge theory with a semi-simple gauge group $G=\prod_i G_i$, the derivatives are taken with respect to the complexified gauge couplings $\tau_i$. In addition, supersymmetric localization provides a powerful tool for computing the four-sphere partition function of these theories \cite{Nekrasov:2002qd, Nekrasov:2003rj,Pestun:2007rz,Okuda:2010ke}. For our purposes, we will only need the tree-level approximation, which yields (see, e.g., \cite{Baume:2020dqd})
\begin{equation}
	Z_{S^4} \sim \prod_i ({\rm Im}\tau_i)^{-{\rm dim}G_i/2} \, .
\end{equation}
As a consequence, the Zamolodchikov metric takes the form of a direct product of hyperbolic upper half-planes 
\begin{equation}
	ds^2 = \sum_i \frac{{\rm dim}G_i}{8\, c} \frac{d\tau_i d \bar \tau_i}{({\rm Im}\tau_i)^2} \, .
\end{equation}
Notice that we have fixed the normalization of the Zamolodchikov metric as in \cite{Perlmutter:2020buo}, so that it corresponds to the moduli-space metric in Planck units in the bulk.

Given this metric, we can now build the orthonormal basis needed to compute the $\vec{\alpha}$-vectors. First, however, let us note that the angular variables $\theta_i$ do not parametrize any infinite-distance limit. For our purposes, we can therefore ignore them. This conveniently leads to a subspace of the conformal manifold on which the Zamolodchikov metric is flat. Parametrized by the gauge couplings $g_i$, it reads
\begin{equation}
	ds^2 = \sum_i \frac{{\rm dim}G_i}{2\, c} \frac{dg_i^2}{g_i^2} \, .
\end{equation}
Consequently, we can define the \emph{flat coordinates} $\hat g_i$ on the conformal manifold.
\begin{equation}
	g_i \sim e^{- \alpha_i \, \hat g_i} \, , \quad \alpha_i=\sqrt{\frac{2\, c}{{\rm dim}G_i}} \, .
\end{equation}
Before proceeding, note that although we used \(\mathcal{N}=2\) for a simpler derivation, this form of the Zamolodchikov metric is expected to hold for any gauge theory at leading order in the \(g_i \to 0\) limit, since it follows from gauge-theoretic tree-level diagrams.

\medskip

Using the flat-coordinate system $\hat g_i$ induced by the Zamolodchikov metric, we can now compute the $\vec{\alpha}$-vectors associated with each tower of HS states. Each gauge sector $G_i$ corresponds to a tower of HS currents whose anomalous dimensions scale as $\gamma_i \sim g_i^2$ and therefore vanish as $g_i \to 0$, so the currents become conserved. Using the AdS/CFT dictionary, this corresponds to a tower of HS states with a mass scale
\begin{equation}
	\frac{M_i}{M_{\rm Pl}} \sim \sqrt{\gamma_i} \sim g_i \sim e^{- \alpha_i \, \hat g_i} \, ,
\end{equation}
where we note that the index $i$ labels distinct gauge factors and their associated towers. From this, we can easily read the $\vec \alpha$-vectors
\begin{equation} \label{eq:alpha-vectors}
	\vec{\alpha}_i = \sqrt{\frac{2\, c}{{\rm dim}G_i}} \, \vec \delta_i = \alpha_{\rm min} \sqrt{\frac{{\rm dim}G}{{\rm dim}G_i}} \, \vec \delta_i \, ,
\end{equation}
where $\vec{\delta}_i$ is the vector whose components are the Kronecker delta, and we used the fact that the $\hat g_i$ are orthonormal coordinates \(
(\nabla \hat g_i)_j \equiv \frac{\partial \hat g_i}{\partial \hat g_j} = \delta_{ij}
\). In the last step, we have factored out $\alpha_{\rm min}$ for later convenience. 

The tower of HS operators discussed above is constructed by acting with $J$ derivatives on a bilinear of two fields transforming in a non-trivial representation of $G_i$ (see, e.g.,~\cite{Skvortsov:2015pea}).\footnote{In general, one must also take the trace to render the operator gauge invariant.} Similarly, one may consider a bilinear of two fields transforming in a non-trivial representation of $G_i \times G_j$, thereby obtaining ``mixed'' towers. Their anomalous dimensions scale as $\gamma_{ij} \sim g_i g_j$. Consequently, they belong to the leading tower only when $g_i \sim g_j$, so that $\gamma_i \sim \gamma_j \sim \gamma_{ij}$. For this reason, they are not relevant for determining the $\alpha$ of the leading tower along any direction, and we may safely ignore them. In particular, they do not affect the shape of the convex hull that we now describe.

\medskip

\begin{figure}[t]
  \centering
    
  \begin{minipage}[t]{0.32\textwidth}
    \centering
    \includegraphics[width=\linewidth]{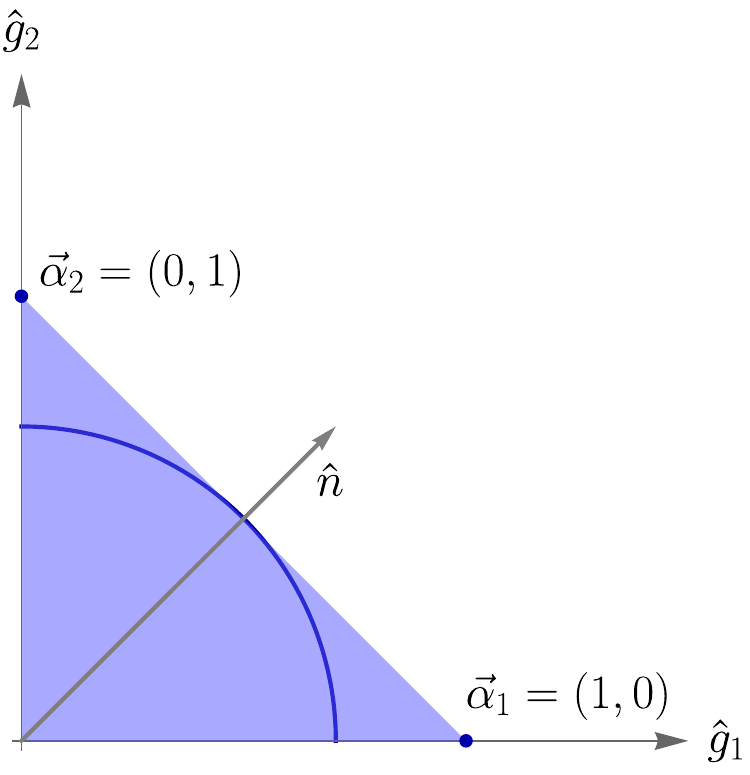}
  \end{minipage}\hfill
  \begin{minipage}[t]{0.32\textwidth}
    \centering
    \includegraphics[width=\linewidth]{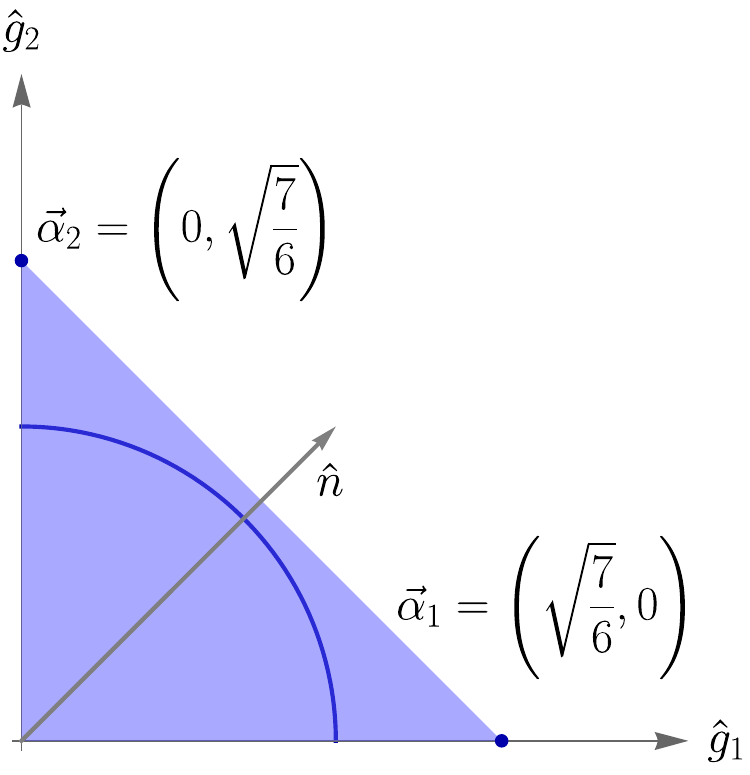}
  \end{minipage}\hfill
  \begin{minipage}[t]{0.32\textwidth}
    \centering
    \includegraphics[width=\linewidth]{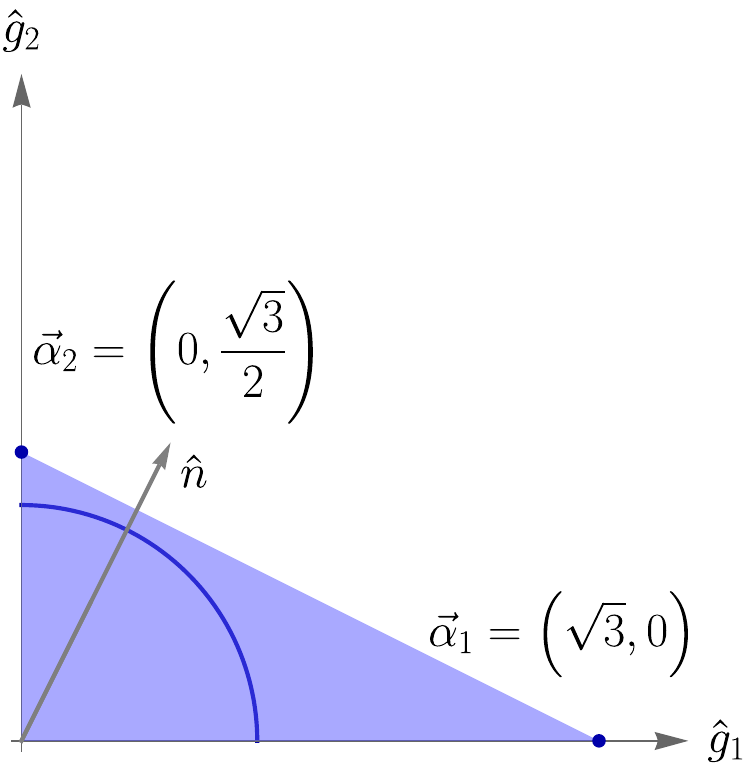}
  \end{minipage}
  \caption{Convex hulls for the weakly-coupled frames of the \(SU(N)\times SU(N)\) circular quiver (left), \(SU(N)\times SU(N)\) linear quiver (middle), and \(SU(N)\times SU(2N)\) circular quiver (right) at large \(N\). The grey arrow depicts the unit normal vector $\hat n$ that points towards the overall-free limit. The curved blue line is the boundary of the ball of radius $1/\sqrt{2}$. The fact that the convex hull contains this ball shows that the bound $\alpha \geq 1/\sqrt{2}$ is satisfied \cite{Calderon-Infante:2020dhm}.}
  \label{fig:convex_hulls_quivers}
\end{figure}

From \eqref{eq:alpha-vectors}, we see that the vectors $\vec{\alpha}_i$ lie on a hyperplane at distance
\begin{equation}
	\alpha_{\rm min} \;=\; \sqrt{\frac{1}{\sum_i 1/\alpha_i^2}}
	\;=\; \sqrt{\frac{2c}{{\rm dim}\,G}},
\end{equation}
from the origin, whose unit normal vector has components
\begin{equation}
	\hat n_i \;=\; \sqrt{\frac{{\rm dim}\,G_i}{{\rm dim}\,G}} \, .
\end{equation}
Thus, the convex hull of the vectors $\vec{\alpha}_i$ (together with the origin) reproduces both the minimum value of $\alpha$ on the conformal manifold and the direction along which it is attained, namely along $g_i \sim g_j$ for any $i$ and $j$. Indeed, one readily checks that
\begin{equation}
    \hat n \cdot \vec{\alpha}_i \;=\; \alpha_{\rm min}\, .
\end{equation}
In Figure \ref{fig:convex_hulls_quivers}, we show some example of these convex hulls. 

As shown above, the convex hull of the vectors $\vec{\alpha}_i$ forms a simplex that covers the entire range of directions $\hat{T}$ for which the computation is valid, namely the weakly coupled frame. This is similar to what has been observed in string compactifications, where this type of structure is called a \emph{frame simplex} \cite{Etheredge:2024tok}.
In fact, the taxonomy program---which aims to classify the possible convex hulls for the Distance Conjecture---starts by identifying all possible frame simplices. This is done via the \emph{taxonomy rules}, which constrain the dot products of the $\vec{\alpha}$-vectors. Our results suggest 
\begin{equation}
\vec{\alpha}_i \cdot \vec{\alpha}_j = 0 \qquad \forall\, i \ne j \, 
\end{equation}
as one of these taxonomy rules in AdS/CFT. In this sense, the types of simplices that can arise in this case seem to be even simpler than those in flat space. A successful characterization of all possible simplices requires classifying all possible lengths of these vectors or, equivalently, the values of $\alpha_{\rm min}$ and $\hat n_i$. Even though this is outside the scope of this work, we would like to provide some comments on this regard. 

\medskip

\paragraph{Populating the allowed range of $\alpha_{\rm min}$.} For any conformal gauge theory, the value of $\alpha_{\min}$ is bounded both above and below due to the conformal collider bounds on $a/c$ \cite{Hofman:2008ar,Hofman:2016awc}. For $\mathcal N=2$ quiver theories in the large-$N$ limit, we have the stronger bounds discussed in Section \ref{ss:bounds}. Nevertheless, it appears that $\alpha_{\rm min}$ can take essentially any value within this interval. To see this explicitly, let us consider the family of $SU(N)^p$ linear quivers. To achieve conformality, we must take $K_1 = K_p = N$ and $K_i = 0$ for any $i\neq 1,p$. The large-$N$ limit of the ratio $a/c$ is given by
\begin{equation} \label{ac-symmetric-quiver}
    \frac{a}{c} = \frac{1+6p}{2+6p} \, .
\end{equation}
By virtue of \eqref{rev:alphamin_ac}, this implies
\begin{equation} \label{alpha-symmetric-quiver}
    \alpha_{\rm min} = \sqrt{\frac{1+3p}{6\,p}} \, .
\end{equation}
We thus see that this class of theories alone suffices to realize infinitely many values of $\alpha_{\rm min}$ within the allowed range. The situation is much more constrained for holographic gauge theories, for which we require $a=c$ at large $N$; hence, the only value allowed in this limit is $\alpha_{\rm min}=1/\sqrt{2}$!

\paragraph{Constraints on the normal vector.}Let us now turn our attention to the normal vector $\hat{n}$. To illustrate the type of constraints it satisfies, let us consider once more $\mathcal{N}=2$ quiver theories with $G_i = SU(N_i)$ in the $N_i \to \infty$ limit. We find that the normal vector points to the direction of the vector $\vec N$, that is,
\begin{equation}
	\hat n = \frac{\vec N}{|\vec N\,|} \, .
\end{equation}
The possible choices of $\vec{N}$ are constrained by the requirement that all of its entries be non-negative integers and by the quiver balancing condition, which imposes 
\begin{equation} \label{eq:n-constraint}
    \vec K = M \cdot \vec N =2 \vec N-A\cdot\vec N \geq 0 \, ,
\end{equation}
where $M$ and $A$ are, respectively, the Cartan and adjacency matrices of the quiver, and we note that the flavor entries must be non-negative. This condition is more constraining for low-dimensional conformal manifolds, for which there are less options for the adjacency matrix $A$. For instance, for $SU(N_1)\otimes SU(N_2)$ quivers we have $A_{12}=A_{21}=n$, with $n$ a positive integer. Equation \eqref{eq:n-constraint} rules out $n\geq 3$. For $n=2$ it fixes $N_1=N_2$, thus recovering the familiar necklace quiver theory. For $n=1$ we have the linear quivers, for which we find 
\begin{equation}
    1 \leq \frac{N_2}{N_1} \leq 2 \, ,
\end{equation}
where the first constraint comes from fixing $N_2 \geq N_1$ without loss of generality. We hence find that the vector $\hat n$ is constrained to lie in the cone defined by the equation above. Nevertheless, the directions $\hat n$ within this range are densely populated by all positive integers $N_1$ and $N_2$. As for the value of $\alpha_{\rm min}$, the situation seems to be much more constrained for holographic theories. For instance, if the quiver is holographic in the large-$N$ limit with $p$ fixed, we showed in Section~\ref{ss:holographic-quivers} that the ranks $N_i$ must be that of an affine ADE quiver up to order-one corrections in the $N_i\to\infty$ limit. In other words, only the vectors $\hat n$ appearing for affine ADE quivers are allowed.

\subsection{Leading Tensionless String vs Hagedorn Temperature} \label{ss:alpha-vs-T_H}

In this section, we compare two pieces of data associated with an infinite-distance, overall weak-coupling limit: the parameter $\alpha$ controlling the exponential behavior in the Distance Conjecture, and the large-$N$ Hagedorn temperature. In Section~\ref{sec:TH}, we argued that the stringy UV completion is determined by the latter. On the other hand, a common lesson from studies of the Distance Conjecture in string compactifications is that the former is tied to the nature of the lightest tower. We therefore ask whether $T_H$ and $\alpha$ track each other in the overall weak-coupling limit of various $\mathcal{N}=2$ quiver CFTs.

A useful benchmark is provided by large-$N$ superconformal gauge theories with \emph{simple gauge group}, for which a one-to-one correspondence between $\alpha$ and $T_H$ was found in \cite{Calderon-Infante:2024oed}. This suggests that in this case only the lightest tower controls the UV resolution of these infinite-distance limits. Indeed, this is always the case for emergent string limits in flat space: the \emph{unique} string whose tension vanishes the fastest determines the UV completion as a weakly coupled string theory \cite{Lee:2019wij,Lee:2019xtm}. In the AdS/CFT setting, however, we will find that several $\mathcal N=2$ quivers can share the same $T_H$ while having different values of $\alpha$, and vice versa. We therefore conclude that either $\alpha$ does not determine the nature of the lightest string, or that the lightest string alone does not capture the UV completion. Unlike in flat space, we will argue that the latter possibility is quite natural in AdS.

\medskip

\textbf{Examples from linear quivers.} We focus on two families of linear $\mathcal N=2$ quivers: the $SU(N)^p$ quivers and the $SU(N)\times\cdots\times SU(pN)$ quivers. In the previous section, we computed the large-$N$ values of the $a/c$ ratio and $\alpha_{\min}$ for the former family (see \eqref{alpha-symmetric-quiver} and \eqref{ac-symmetric-quiver}). The latter theory is conformal when $K_p=(p+1)N$ and $K_i=0$ for all $i\neq p$. Using these results, we obtain
\begin{equation}
\frac{a}{c}=\frac{3+4p}{4+4p}\,,\qquad 
\alpha_{\min}=\sqrt{\frac{1+p}{1+2p}}\,.
\end{equation}
For any fixed length $p>1$, the two families yield different values of $\alpha_{\min}$. By contrast, as shown in Section \ref{sec:TH}, the Hagedorn temperature $T_H$ for linear quivers depends only on $p$. Therefore, in the overall weak-coupling limit, these provide examples with the same $T_H$ but different $\alpha_{\min}$. Conversely, one can also find theories with the same $\alpha_{\min}$ but different $T_H$. For example, this occurs for the $SU(N)^{p_1}$ and $SU(N)\times\cdots\times SU(p_2N)$ quivers when $p_1=1+2k$ and $p_2=1+3k$, for any integer $k$. Altogether, we conclude that for $\mathcal N=2$ quivers there is neither a one-to-one nor a one-to-many relation between $\alpha_{\min}$ and $T_H$.

\medskip

\paragraph{Implications.} To see the implications of this result more clearly, let us recall the conditions that must be satisfied for the values of $\alpha$ and $T_H$ to necessarily coincide along an infinite-distance limit:
\begin{itemize}
    \item[(i)] The parameter $\alpha$ identifies the type of the leading tensionless string.
    \item[(ii)] The Hagedorn density of states is dominated by this leading tensionless string.
\end{itemize}
In our examples it is not straightforward to determine which of the two conditions fails, since it is unclear how to identify the leading tensionless string from the CFT spectrum in the infinite-distance limit. Instead, we argue that condition~(ii) should not be expected to hold generically in AdS, in contrast to flat space.

In flat space, all string excitations become massless in the tensionless limit. The resulting spectrum exhibits Hagedorn growth, with a Hagedorn temperature set parametrically by the characteristic scale of the tower, \(T_H \sim M_s\). Therefore, since a \emph{unique} string becomes tensionless at the fastest rate in the infinite-distance limit, condition (ii) is satisfied.

In AdS/CFT, the situation is qualitatively different. Our gauge-theory analysis shows that in the overall weak-coupling limit the Hagedorn temperature does \emph{not} tend to zero; instead, it remains of order the AdS scale. Equivalently, $T_H$ decouples from the mass scale of the higher-spin (HS) tower that becomes massless in AdS (or Planck) units. From the bulk viewpoint, this indicates that not all string excitation modes become parametrically light in the limit. Indeed, as discussed further in Appendix~\ref{App: HS Density}, the massless HS tower is not exponentially dense and therefore does not control the Hagedorn temperature. Rather, $T_H$ is set by a different sector of excitations whose masses remain of order the AdS scale in the limit.

This makes it natural for (ii) to fail as follows. Consider an infinite-distance limit in which one string becomes tensionless, while another extended sector remains at the AdS scale (or becomes light only at a subleading rate). Then $T_H$ can be sensitive to both sectors, and the Hagedorn growth need not be dominated by the leading tensionless string. In AdS, unlike flat space, $T_H$ can therefore receive contributions beyond the lightest tower.

In fact, condition (ii) can be satisfied in AdS only in a tensionless-string limit in which all other strings---and, more generally, all extended objects---become infinitely heavy compared to the AdS scale. This occurs naturally in the $g_s \to 0$ limit of Type IIB string theory on a fixed background: all D$p$-branes and the NS5-brane---possibly wrapping cycles of the fixed internal geometry---then become infinitely heavy. Accordingly, as shown in Section~\ref{ss:holographic-quivers}, the one-to-one correspondence between $\alpha_{\min}$ (or $a/c$) and $T_H$ is restored for holographic quiver CFTs whose bulk dual is precisely of this type.

\medskip

In conclusion, assuming condition (i) holds, the absence of a one-to-one correspondence between $\alpha$ and $T_H$ indicates that the bulk description of the overall weak-coupling limit of general $\mathcal N=2$ quivers involves extended objects that do not become infinitely heavy relative to the AdS scale. Consequently, the UV completion (as encoded by $T_H$) is not determined by the leading tower. An analogous phenomenon occurs in flat space, even though it arises in decompactification limits rather than tensionless-string limits. A simple example is a two-torus whose radii are sent to infinity at different rates: even if the leading tower is the KK tower associated with one radius, the full UV completion is sensitive to both internal directions. In this sense, tensionless-string limits in AdS resemble decompactification limits more than tensionless-string limits in flat space.

\section{Conclusions} \label{sec:conclusion}
In this work, we present a detailed study of the CFT Distance Conjecture for four-dimensional $\mathcal{N}=2$ superconformal quiver gauge theories. Emphasizing the interpretation of the overall free limit as a tensionless-string limit in AdS, we investigate the role of the large-$N$ Hagedorn temperature as a coarse probe of the bulk stringy UV completion and derive bounds on the order-one constant that controls the exponential decay in the Distance Conjecture.

In Section~\ref{sec:TH}, we computed the Hagedorn temperature from the large-$N$ thermal partition function on $S^3 \times S^1$ at the overall free point for several types of $\mathcal{N}=2$ $SU$ quivers with hypermultiplets in the bifundamental and fundamental representations. For \emph{linear quivers}, we found a striking universality: the Hagedorn temperature $T_H$ is insensitive to the detailed rank and flavor data and depends only on the quiver length. We then interpreted this result in the Hanany--Witten realization, where the length of the quiver is tied to the number of NS5-branes and therefore to the type of string theory in which the bulk is embedded. Turning to \emph{holographic quivers}, we showed that $T_H$ coincides with that of $\mathcal{N}=4$ SYM, independently of the detailed quiver data, matching the expectation that these theories are UV-completed by the fundamental Type~IIB string.

In Section~\ref{sec: alpha}, we analyzed the exponential rate~$\alpha$ governing how the conformal dimensions of higher-spin (HS) currents approach the unitarity bound. Along the overall-free direction, we derived strong two-sided bounds on the minimal rate,
\(
\frac{1}{\sqrt{2}} \le \alpha_{\min} \le \sqrt{\frac{2}{3}},
\)
in the large-$N$ regime. We also proved the universal lower bound $\alpha \ge \frac{1}{\sqrt{2}}$ proposed in~\cite{Perlmutter:2020buo}, including at finite~$N$. We then moved beyond the overall-free ray by computing the vectors~$\vec{\alpha}$ that encode the exponential decay rates of the HS towers along any (partial) weakly coupled limit. The convex hull of these vectors takes the form of a \emph{frame simplex} and is completely characterized by the value of $\alpha_{\rm min}$ and a unit normal vector~$\hat n$ that encodes how the overall-free direction is selected within a given weakly coupled frame.
Finally, we clarified the interplay between $\alpha$ and $T_H$ in the overall-free limit, which are expected to encode the nature of the tower of light states and the stringy UV completion in the bulk, respectively. While a one-to-one correspondence holds for simple gauge groups~\cite{Calderon-Infante:2024oed}, this feature is lost for general non-holographic quivers. We argued that this is not necessarily in disagreement with the aforementioned interpretations of $\alpha$ and $T_H$. The reason is that in AdS, unlike in flat space, the Hagedorn growth is not determined solely by the leading tensionless string unless all other extended objects decouple from the AdS scale in the infinite-distance limit.

\paragraph{Future directions.}
Our results leaves several interesting questions for future research, including a variety of generalizations beyond the class of $\mathcal{N}=2$ $SU$ quiver gauge theories with fundamental and bifundamental hypermultiplets considered here. 

\begin{itemize}
  \item It would be interesting to test the relation between $T_H$ and the stringy UV completion beyond $\mathcal{N}=2$ linear $SU$ quivers with hypermultiplets in fundamental and bifundamental representations. It would be particularly valuable to study models with known realizations in terms of Hanany--Witten brane configurations, since these provide a concrete expectation for the type of stringy UV-completion and hence for the Hagedorn temperature that can then be computed from the CFT. For instance, orientifolds of the basic setup studied in Section~\ref{ss:linear quivers} lead to linear quivers involving not only $SU$ but also $SO$ and $USp$ gauge groups, as well as hypermultiplets in (anti)symmetric representations. Furthermore, orbifolds can be used to go beyond linear quivers.

  \item One can extend the analysis of holographic $\mathcal{N}=2$ quivers (studied in Section \ref{ss:holographic-quivers}) beyond $SU$ gauge groups, and also include hypermultiplets in the (anti)symmetric representations. It would be interesting to determine the consequences of imposing the balancing condition and $a \simeq c$ for these models, and whether they lead to theories with Type~IIB bulk duals and the Hagedorn temperature of $\mathcal{N}=4$ SYM.

  \item It is worthwhile to generalize our derivation of bounds on $\alpha_{\rm min}$ to $\mathcal{N}=2$ quivers beyond $SU$ gauge groups, including hypermultiplets in the (anti)symmetric representations. The key step is to write down the balancing condition for these quivers and try to incorporate it naturally into the expression for $n_h/n_v$.

  \item There has recently been renewed interest in how duality symmetries inform our understanding of a consistent theory of quantum gravity (see, e.g.,~\cite{Delgado:2024skw,Chen:2025rkb,Mohseni:2025tig,Fraiman:2025yrx,Baykara:2025gcc}). Part of this duality group is encoded in how different simplices of the convex hull are glued together~(see e.g. \cite{Etheredge:2022opl,Calderon-Infante:2023ler,Etheredge:2023odp,Castellano:2023jjt,Castellano:2023stg,Etheredge:2024tok,Grieco:2025bjy,Aoufia:2025ppe}). It would be interesting to consider 4d $\mathcal{N}=2$ SCFTs whose distinct S-duality frames on the conformal manifold are known, compute the $\vec{\alpha}$-vectors for the HS towers across these frames, and study how the corresponding frame simplices are glued together. A promising arena for these studies is that of 4d $\mathcal{N}=2$ SCFTs with class~S realizations ~\cite{Gaiotto:2009we}.

  \item Another interesting direction is to further develop a taxonomy program for asymptotic regimes in AdS/CFT. It would be interesting to study constraints not only on $\alpha_{\rm min}$ but also on the unit normal vector $\hat n$ introduced in Section~\ref{ss:convex-hulls}. This could potentially lead to a classification of simplices describing (partially) weakly coupled frames. Furthermore, studying the possible ways in which these simplices can be glued together might offer insights into S-dualities on the conformal manifold of these theories.

  \item When computing the $\vec{\alpha}$-vectors in Section~\ref{ss:convex-hulls}, we considered only motion along the conformal manifold. For the study of the Distance Conjecture in AdS/CFT, it is also interesting to include an additional direction along which the central charge of the CFT increases. Via AdS/CFT, this corresponds to taking the AdS length to infinity in Planck units, which has been conjectured to lead to a tower of light states \cite{Lust:2019zwm}. In concrete examples, this occurs when a family of AdS vacua approaches an infinite-distance limit in field space, leading to the decompactification of extra dimensions. This central-charge direction was added in \cite{Calderon-Infante:2024oed} for holographic gauge theories with simple gauge group via their bulk duals. It would be interesting to study how to incorporate this direction for non-holographic theories directly in the CFT, and to explore how it completes the convex hull of four-dimensional SCFTs with multi-dimensional conformal manifolds.

  \item It would be interesting to develop techniques for classifying operators in a CFT according to whether they plausibly originate from distinct string sectors. Among other things, this could shed light on the discussion of the leading tensionless string versus the UV completion in Section~\ref{ss:alpha-vs-T_H}.

\end{itemize}

This work has taken an initial step toward improving our understanding of the CFT Distance Conjecture for theories with (complex) multi-dimensional conformal manifolds, and of the bulk physics corresponding to higher-spin infinite-distance limits on the CFT side. We hope our results encourage new and exciting studies of asymptotic regimes in AdS/CFT. Regarding the bulk interpretation as tensionless-string limits, we hope that the ideas explored in this paper help pave the way toward eventually constructing string-theoretic duals for non-holographic CFTs, which are of great physical importance because they may help us understand real-world phenomena---such as QCD confinement---from a gravitational dual perspective.

\section*{Acknowledgments} 
We are especially grateful to Angel M. Uranga and Irene Valenzuela for illuminating discussions and useful feedback. J.C. would also like to thank them for collaborations on related topics. We also thank Florent Baume, Craig Lawrie, Leonardo Rastelli and Jaewon Song for useful conversations. The work of J.C. is supported by the Walter Burke Institute for Theoretical Physics at Caltech, by the U.S. Department of Energy, Office of Science, Office of High Energy Physics, under Award Number DE-SC0011632, and was performed in part at Aspen Center for Physics, which is supported by National Science Foundation grant PHY-2210452. The work of A.M. is supported in part by a grant from the Simons Foundation (602883, CV) and the DellaPietra Foundation. J.C. is also grateful to Marta Igarzabal Galdos for her continuous encouragement and support.

\appendix

\section{Explicit Calculation of the Hagedorn Temperature for ADE Quivers} \label{App:ADE explicit}
In this appendix, we present explicit calculations of the Hagedorn temperature for affine ADE quivers by computing the determinant of the matrices $M^{(l)}$ that appear in the effective action for the large-$N$ partition function as in \eqref{eq:Seff-matrix}. 

\subsection{Type D Quivers}
\begin{figure}[h]
  \centering
  \includegraphics[width=0.8\textwidth]{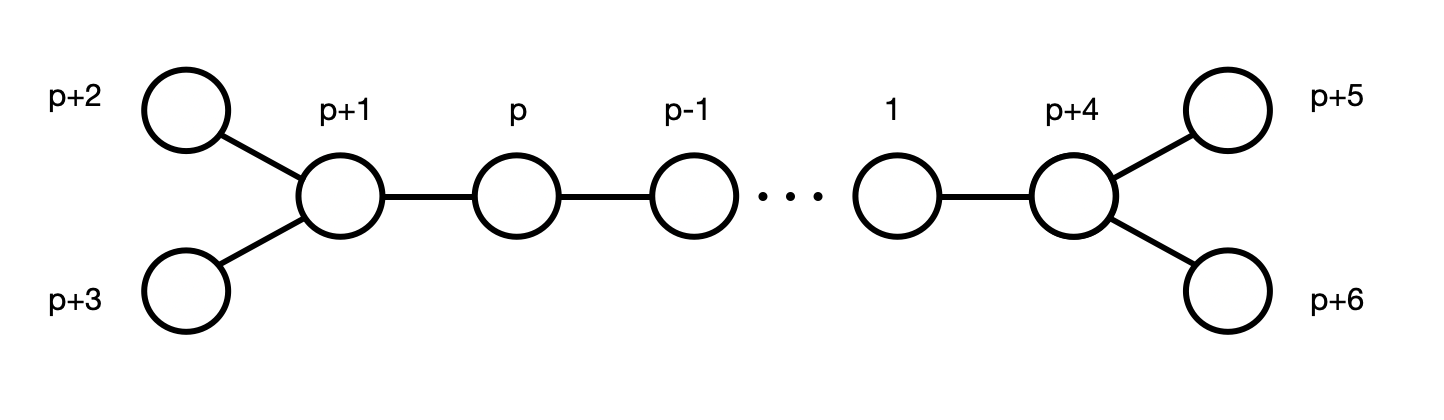}
  \caption{A quiver of type $D$.}
  \label{fig:Dquiver}
\end{figure}

Consider an affine quiver of type $D$ in the ADE classification with gauge group
\begin{equation}
G \;=\; \bigotimes_{i=1}^{p+6} SU(N_i),
\end{equation}
where, for convenience, the nodes are labeled as in Figure~\ref{fig:Dquiver}.
The ranks are $N$ for the four outer nodes
and $2N$ for the internal ones. This will however be irrelevant for the derivation as long as we work in the $N_i \to \infty$ limit.

Proceeding as for the linear quivers in Section \ref{sss:linear}, the large-$N$ thermal partition function can be written as an integral over the Fourier modes $\rho_i^{(l)}$ of the eigenvalue distributions
\begin{equation}
Z(x)\;=\;\prod_{k=1}^{p}\prod_{l=1}^{\infty}\frac1{l\pi}
   \int d^{2}\rho^{(l)}_{k}\,e^{-S_{\text{eff}}} \, .
\end{equation}
Up to linear and constant terms, which are irrelevant for determining the Hagedorn temperature, the effective action reads
\begin{equation}
S_{\mathrm{eff}}=\bar\rho^{(l)}\!\cdot M^{(l)}\!\cdot \rho^{(l)} \, ,
\end{equation}
For convenience, we define
\begin{equation}\label{xieta}
\xi^{(l)} \equiv 1-z_{\mathrm{adj}}(x,l),\qquad
\eta^{(l)} \equiv -\,z_{\mathrm{biF}}(x,l).
\end{equation}
Then, $M^{(l)}$ is the tridiagonal Toeplitz matrix deformed by the two branchings at the ends:
\begin{equation}\label{typeDmatrix}
M^{(l)}=
\begin{bmatrix}
  \xi^{(l)} & 0 & \eta^{(l)} & 0 & 0 & \cdots & 0 & 0 & 0 & 0 \\
  0 & \xi^{(l)} & \eta^{(l)} & 0 & 0 & \cdots & 0 & 0 & 0 & 0 \\
  \eta^{(l)} & \eta^{(l)} & \xi^{(l)} & \eta^{(l)} & 0
      & \cdots & 0 & 0 & 0 & 0 \\
  0 & 0 & \eta^{(l)} & \xi^{(l)} & \eta^{(l)}
      & \cdots & 0 & 0 & 0 & 0 \\
  0 & 0 & 0 & \eta^{(l)} & \xi^{(l)}
      & \cdots & 0 & 0 & 0 & 0 \\
  \vdots & \vdots & \vdots & \vdots & \vdots & \ddots & \eta^{(l)} & 0 & 0 & 0 \\
  0 & 0 & 0 & 0 & \cdots & \eta^{(l)} &
      \xi^{(l)} & \eta^{(l)} & 0 & 0 \\
  0 & 0 & 0 & 0 & 0 & \cdots & \eta^{(l)} &
      \xi^{(l)} & \eta^{(l)} & \eta^{(l)} \\
  0 & 0 & 0 & 0 & 0 & \cdots & 0 &
      \eta^{(l)} & \xi^{(l)} & 0 \\
  0 & 0 & 0 & 0 & 0 & \cdots & 0 &
      \eta^{(l)} & 0 & \xi^{(l)}
\end{bmatrix}_{(p+6)\times(p+6)} .
\end{equation}
As explained in \ref{sss:linear}, the Hagedorn divergence happens for the smallest temperature such that the determinant of any of the matrices above vanishes.

To compute the determinant of the matrix in \eqref{typeDmatrix}, as an intermediate step, consider the matrix that has a single branch at one end
\begin{equation}
m^{(l)}=
\begin{bmatrix}
  \xi^{(l)} & 0 & \eta^{(l)} & 0 & \cdots & 0 \\
  0 & \xi^{(l)} & \eta^{(l)} & 0 & \cdots & 0 \\
  \eta^{(l)} & \eta^{(l)} & \xi^{(l)} & \eta^{(l)} & \cdots & 0 \\
  0 & 0 & \eta^{(l)} & \xi^{(l)} & \cdots & 0 \\
  \vdots & \vdots & \vdots & \vdots & \ddots & \eta^{(l)} \\
  0 & 0 & 0 & 0 & \eta^{(l)} & \xi^{(l)}
\end{bmatrix}_{(p+3)\times(p+3)}.
\end{equation}

\paragraph{Schur complement formula.}
For a block matrix
\[
M=\begin{bmatrix}A & B\\ C & D\end{bmatrix},
\]
if $A$ and $D$ are invertible then
\begin{equation}
\det M = \det A\,\det\!\bigl(D-CA^{-1}B\bigr)
       = \det D\,\det\!\bigl(A-BD^{-1}C\bigr).
\end{equation}

\subsection*{Step 1: determinant of the one-branch matrix $m^{(l)}$}
Let us write $m^{(l)}$ in block form above with
\begin{equation}
A=
\begin{bmatrix}
\xi^{(l)} & 0 & \eta^{(l)}\\
0 & \xi^{(l)} & \eta^{(l)}\\
\eta^{(l)} & \eta^{(l)} & \xi^{(l)}
\end{bmatrix},
\qquad
D=
\begin{bmatrix}
\xi^{(l)} & \eta^{(l)} & 0 & \cdots & 0\\
\eta^{(l)} & \xi^{(l)} & \eta^{(l)} & \ddots & \vdots\\
0 & \eta^{(l)} & \xi^{(l)} & \ddots & 0\\
\vdots & \ddots & \ddots & \ddots & \eta^{(l)}\\
0 & \cdots & 0 & \eta^{(l)} & \xi^{(l)}
\end{bmatrix}_{p\times p},
\end{equation}
and
\begin{equation}
B=C^{T}=
\begin{bmatrix}
0 & 0 & \cdots & 0\\
0 & 0 & \cdots & 0\\
\eta^{(l)} & 0 & \cdots & 0
\end{bmatrix}_{3\times p}.
\end{equation}
A direct computation yields
\begin{equation}\label{app:CAinvB}
C\,A^{-1}B
=\gamma^{(l)}
\begin{bmatrix}
1 & 0\\
0 & 0
\end{bmatrix}_{p\times p},
\qquad
\gamma^{(l)}=
\frac{\bigl(1-z_{\mathrm{adj}}(x,l)\bigr)\,z_{\mathrm{biF}}(x,l)^2}{
\bigl(1-z_{\mathrm{adj}}(x,l)\bigr)^2-2\,z_{\mathrm{biF}}(x,l)^2}.
\end{equation}
Hence, we have
\begin{equation}
\det m^{(l)}
=\det A\;\det\!\bigl(D-C A^{-1}B\bigr)
=\xi^{(l)}\Big((\xi^{(l)})^2-2(\eta^{(l)})^2\Big)\;\bigl(d_{p}^{(l)}-\gamma^{(l)} d_{p-1}^{(l)}\bigr),
\end{equation}
where $d_n^{(l)}$ denotes the determinant of the $n\times n$ tridiagonal Toeplitz block with
diagonal entry $\xi^{(l)}$ and off-diagonal entry $\eta^{(l)}$, so that
\begin{equation}\label{recursion relation}
d_0^{(l)}=1,\qquad d_1^{(l)}=\xi^{(l)},\qquad
d_n^{(l)} = \xi^{(l)}\,d_{n-1}^{(l)} - \bigl(\eta^{(l)}\bigr)^2 d_{n-2}^{(l)}\quad(n\ge2),
\end{equation}
with the closed form
\begin{equation}\label{explicitreddet}
d_n^{(l)} = \frac{\bigl|\eta^{(l)}\bigr|^n \sin\!\bigl((n+1)\theta\bigr)}{\sin\theta},
\qquad
\cos\theta \equiv \frac{\xi^{(l)}}{2|\eta^{(l)}|}.
\end{equation}

\subsection*{Step 2: determinant of the two-branch matrix $M^{(l)}$}
\begin{figure}[h]
    \centering
    \includegraphics[width=14cm]{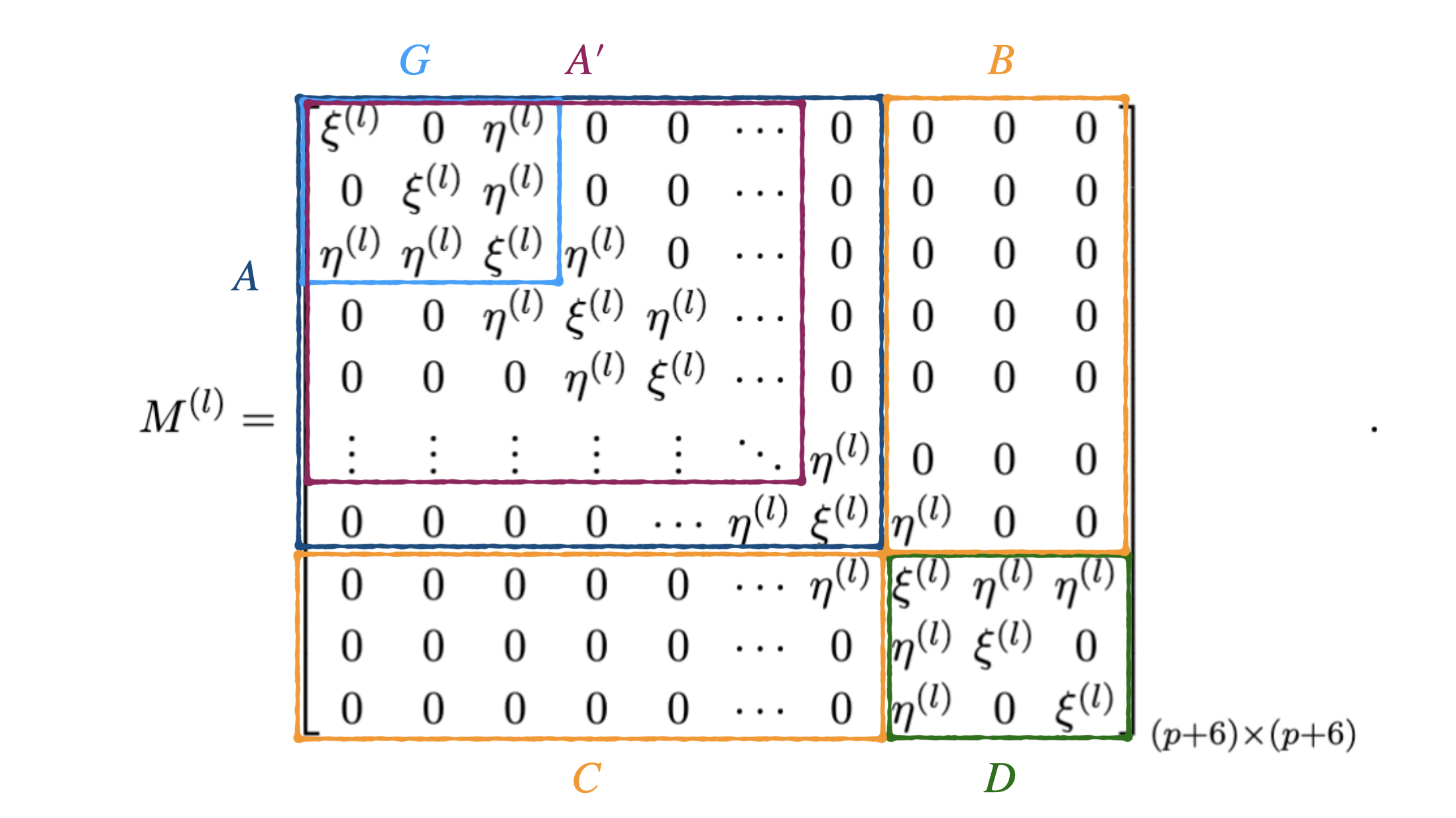}
    \caption{Applying the Schur complement formula.}
    \label{schur2}
\end{figure}

Having computed the determinant with a single branching, we now apply the Schur
formula to the original matrix
(see Figure \ref{schur2}; note that the blocks \(A,B,C,D\) are now re-defined).
When the corresponding tridiagonal block $D$ is invertible, we may write
\begin{equation}
\det M^{(l)}=\det D\;\det\!\bigl(A-BD^{-1}C\bigr).
\end{equation}
Thus, $\det M^{(l)}=0$ implies $\det\!\left(A-BD^{-1}C\right)=0$.

One finds
\begin{equation}
BD^{-1}C
=\gamma^{(l)}
\begin{bmatrix}
0 & 0\\
0 & 1
\end{bmatrix}_{(p+3)\times(p+3)},
\end{equation}
with the same $\gamma^{(l)}$ as in \eqref{app:CAinvB}. Writing $A'$ for the minor of $A$ appropriate to
this Schur reduction, the condition $\det(A-BD^{-1}C)=0$ becomes
\begin{equation}
\det A - \gamma^{(l)}\det A' = 0
\qquad\Longrightarrow\qquad
\gamma^{(l)}=\frac{\det A}{\det A'}.
\end{equation}
Using the expression for $\det m^{(l)}$ obtained above for $\det A$ and $\det A'$, we find that the zeros of the determinant determining the Hagedorn temperature are given by
\begin{equation}\label{Dgammaeq}
\gamma^{(l)}
=\frac{d_{p}^{(l)}-\gamma^{(l)}\,d_{p-1}^{(l)}}{d_{p-1}^{(l)}-\gamma^{(l)}\,d_{p-2}^{(l)}}\,.
\end{equation}

Substituting \eqref{explicitreddet} and \eqref{app:CAinvB} into \eqref{Dgammaeq}, one arrives at the condition
\begin{align}
z_{\mathrm{biF}}^{p}\,\csc\theta\;\Bigl[
  &(z_{\mathrm{adj}}-1)^{2}\,z_{\mathrm{biF}}^{2}\,
      \sin\!\bigl((p-1)\theta\bigr)\notag\\
  &+\bigl((1-z_{\mathrm{adj}})^2-2z_{\mathrm{biF}}^{2}\bigr)
   \Bigl(
     2(z_{\mathrm{adj}}-1)z_{\mathrm{biF}}\,\sin(p\theta)
     +\bigl((1-z_{\mathrm{adj}})^2-2z_{\mathrm{biF}}^{2}\bigr)
       \sin\!\bigl((p+1)\theta\bigr)
   \Bigr)
\Bigr] \;=\;0.
\end{align}
Finally, rewriting in terms of the single-particle partition functions for the vector and chiral multiplets,
\(z_{\text{adj}} = z_v + z_c\) and \(z_{\text{biF}} = z_c\), we find
\begin{align}
z_{c}^{\,p}\,\csc\theta
\Bigl[
  &\frac{
    z_{c}^{2}\,(z_{c}+z_{v}-1)^{2}\,
    \sin\!\bigl((p-1)\theta\bigr)
  }{
    z_{c}^{2}-2z_{c}(z_{v}-1)-(z_{v}-1)^{2}
  }
  -2\,z_{c}\,(z_{c}+z_{v}-1)\,\sin(p\theta)\notag\\
  &\hspace{2.5cm}
  -\bigl(-z_{c}^{2}+2z_{c}(z_{v}-1)+(z_{v}-1)^{2}\bigr)\,
  \sin\!\bigl((p+1)\theta\bigr)
\Bigr] \;=\;0.
\label{Dzeroequ}
\end{align}

For 
\(
z_v + 3 z_c = 1,
\)
namely the equation that determines the Hagedorn temperature of \(\mathcal N = 4\) SYM \cite{Calderon-Infante:2024oed}, the condition reduces to
\begin{equation}
z_{c}^{\,p}\,\sin(p\theta)\,\tan\!\bigl(\tfrac{\theta}{2}\bigr) = 0 
\qquad \text{for } \cos\theta = 1 ,
\end{equation}
which is indeed satisfied. Plugging \(z_v = 1 - a\,z_c\) into \eqref{explicitreddet}, we find
\begin{align}
&\cos\theta = \frac{a-1}{2}.
\end{align}
We immediately see that \(-1 \leq a \leq 3\). Hence, the smallest temperature (corresponding to the largest value of \(a\)) is obtained for \(a = 3\), which coincides with the Hagedorn temperature of \(\mathcal N = 4\) SYM.

\subsection{Type E Quivers}
\begin{figure}[h]
  \centering
  \includegraphics[width=\textwidth]{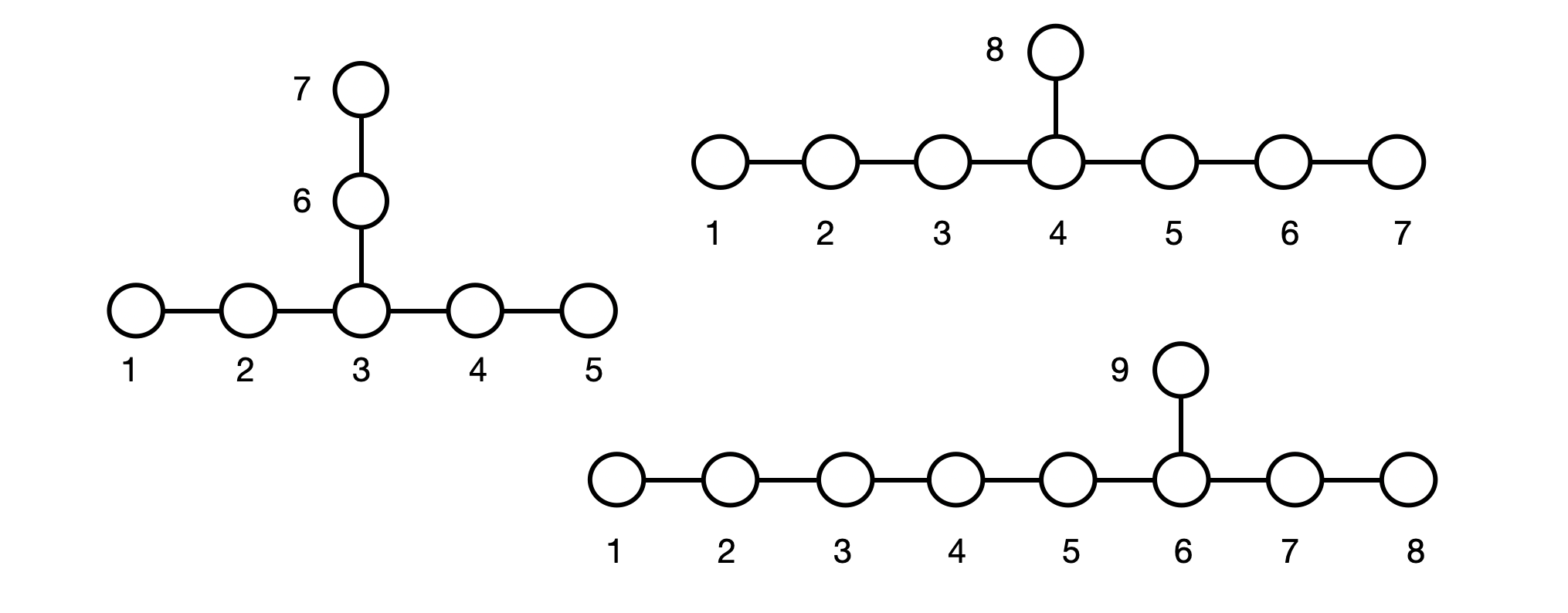}
  \caption{$E_6$, $E_7$, and $E_8$ quivers.}
  \label{fig:Equivers}
\end{figure}

Consider the quivers of type $E$ in the ADE classification, shown in Figure~\ref{fig:Equivers}.
The computation of the large-$N$ partition function again proceeds as in the linear quiver case in Section \ref{sss:linear}. We will specify the ranks $N_i$ for each case, but we note again that this will not affect the result for the Hagedorn temperture $T_H$, which is solely determined by the structure of the interaction matrix between the eigenvalue distributions, $M^{(l)}$.

\subsubsection*{$E_6$ Quiver}
Consider the $E_6$ quiver with node ranks, listed in graph-label order, as
\[
(N,\,2N,\,3N,\,2N,\,N,\,2N,\,N).
\]
Then, the interaction matrix is
\begin{equation}
M_{E_6}^{(l)}=
\begin{bmatrix}
\xi^{(l)} & \eta^{(l)} & 0 & 0 & 0 & 0 & 0 \\
\eta^{(l)} & \xi^{(l)} & \eta^{(l)} & 0 & 0 & 0 & 0 \\
0 & \eta^{(l)} & \xi^{(l)} & \eta^{(l)} & 0 & \eta^{(l)} & 0 \\
0 & 0 & \eta^{(l)} & \xi^{(l)} & \eta^{(l)} & 0 & 0 \\
0 & 0 & 0 & \eta^{(l)} & \xi^{(l)} & 0 & 0 \\
0 & 0 & \eta^{(l)} & 0 & 0 & \xi^{(l)} & \eta^{(l)} \\
0 & 0 & 0 & 0 & 0 & \eta^{(l)} & \xi^{(l)}
\end{bmatrix}.
\end{equation}
The Hagedorn temperature is determined by $\det(M_{E_6}^{(l)})=0$, namely
\begin{align}
\det(M_{E_6}^{(l)})
&=(1-z_{\mathrm{adj}})\Bigl((1-z_{\mathrm{adj}})^2-4z_{\mathrm{biF}}^2\Bigr)\Bigl((1-z_{\mathrm{adj}})^2-z_{\mathrm{biF}}^2\Bigr)^2 \notag\\
&=(1-z_v)^2(1-z_c-z_v)(1-2z_c-z_v)^2\Bigl(1-4z_c^2-2(z_c+z_v)+(z_c+z_v)^2\Bigr)=0.
\end{align}
The solutions can be written as $z_v+a\,z_c=1$ with $a\in\{-1,0,1,2,3\}$. Hence the smallest temperature
is again set by $a=3$, i.e.\ $z_v+3z_c=1$, which coincides with the Hagedorn temperature of $\mathcal N=4$ SYM.

\subsubsection*{$E_7$ Quiver}
For the $E_7$ quiver with ranks $(N,2N,3N,4N,3N,2N,N,2N)$, ordered according to the labels on the graph, we have 
\begin{equation}
M_{E_7}^{(l)}=
\begin{bmatrix}
\xi^{(l)} & \eta^{(l)} & 0 & 0 & 0 & 0 & 0 & 0 \\
\eta^{(l)} & \xi^{(l)} & \eta^{(l)} & 0 & 0 & 0 & 0 & 0 \\
0 & \eta^{(l)} & \xi^{(l)} & \eta^{(l)} & 0 & 0 & 0 & 0 \\
0 & 0 & \eta^{(l)} & \xi^{(l)} & \eta^{(l)} & 0 & 0 & \eta^{(l)} \\
0 & 0 & 0 & \eta^{(l)} & \xi^{(l)} & \eta^{(l)} & 0 & 0 \\
0 & 0 & 0 & 0 & \eta^{(l)} & \xi^{(l)} & \eta^{(l)} & 0 \\
0 & 0 & 0 & 0 & 0 & \eta^{(l)} & \xi^{(l)} & 0 \\
0 & 0 & 0 & \eta^{(l)} & 0 & 0 & 0 & \xi^{(l)}
\end{bmatrix}.
\end{equation}
The Hagedorn condition $\det(M_{E_7}^{(l)})=0$ can be expressed as
\begin{align}
\det(M_{E_7}^{(l)})
&=(z_{\mathrm{adj}}-1)^2(z_{\mathrm{adj}}-2z_{\mathrm{biF}}-1)(z_{\mathrm{adj}}-z_{\mathrm{biF}}-1)
(z_{\mathrm{adj}}+z_{\mathrm{biF}}-1)(z_{\mathrm{adj}}+2z_{\mathrm{biF}}-1)\notag\\
&\hspace{1.8cm}\times\Bigl((z_{\mathrm{adj}}-1)^2-2z_{\mathrm{biF}}^2\Bigr) \notag\\
&=\Bigl(6 z_{c}^{5} - 13 z_{c}^{4}(z_{v}-1) - 8 z_{c}^{3}(z_{v}-1)^{2}
+ 8 z_{c}^{2}(z_{v}-1)^{3} + 6 z_{c}(z_{v}-1)^{4} + (z_{v}-1)^{5}\Bigr)\notag\\
&\hspace{1.8cm}\times (z_{v}-1)\,(z_{c}+z_{v}-1)^{2}=0.
\end{align}
Equivalently, $z_v+a\,z_c=1$ with $a\in\{-1,1-\sqrt2,0,1,2,1+\sqrt2,3\}$, and the smallest temperature
again corresponds to $a=3$, coinciding with the Hagedorn temperature of $\mathcal N=4$ SYM.

\subsubsection*{$E_8$ Quiver}
For the $E_8$ quiver with ranks $(N, 2N, 3N, 4N, 5N, 6N, 4N, 2N, 3N)$, ordered according to the node labels in the diagram, we have
\begin{equation}
M_{E_8}^{(l)}=
\begin{bmatrix}
\xi^{(l)} & \eta^{(l)} & 0 & 0 & 0 & 0 & 0 & 0 & 0 \\
\eta^{(l)} & \xi^{(l)} & \eta^{(l)} & 0 & 0 & 0 & 0 & 0 & 0 \\
0 & \eta^{(l)} & \xi^{(l)} & \eta^{(l)} & 0 & 0 & 0 & 0 & 0 \\
0 & 0 & \eta^{(l)} & \xi^{(l)} & \eta^{(l)} & 0 & 0 & 0 & 0 \\
0 & 0 & 0 & \eta^{(l)} & \xi^{(l)} & \eta^{(l)} & 0 & 0 & 0 \\
0 & 0 & 0 & 0 & \eta^{(l)} & \xi^{(l)} & \eta^{(l)} & 0 & \eta^{(l)} \\
0 & 0 & 0 & 0 & 0 & \eta^{(l)} & \xi^{(l)} & \eta^{(l)} & 0 \\
0 & 0 & 0 & 0 & 0 & 0 & \eta^{(l)} & \xi^{(l)} & 0 \\
0 & 0 & 0 & 0 & 0 & \eta^{(l)} & 0 & 0 & \xi^{(l)}
\end{bmatrix}.
\end{equation}
The Hagedorn condition $\det(M_{E_8}^{(l)})=0$ factorizes as
\begin{align}
\det(M_{E_8}^{(l)})
&=-(z_{\mathrm{adj}}-1)(z_{\mathrm{adj}}-2z_{\mathrm{biF}}-1)(z_{\mathrm{adj}}-z_{\mathrm{biF}}-1)(z_{\mathrm{adj}}+z_{\mathrm{biF}}-1)(z_{\mathrm{adj}}+2z_{\mathrm{biF}}-1)\notag\\
&\quad\times\Bigl(z_{\mathrm{adj}}^2+z_{\mathrm{adj}}(z_{\mathrm{biF}}-2)-z_{\mathrm{biF}}(z_{\mathrm{biF}}+1)+1\Bigr)
\Bigl(z_{\mathrm{adj}}^2-z_{\mathrm{adj}}(z_{\mathrm{biF}}+2)-z_{\mathrm{biF}}^2+z_{\mathrm{biF}}+1\Bigr)\notag\\
&=-\Bigl(6 z_{c}^{7} + 11 z_{c}^{6}(z_{v}-1) - 24 z_{c}^{5}(z_{v}-1)^{2} - 30 z_{c}^{4}(z_{v}-1)^{3}
+ 8 z_{c}^{3}(z_{v}-1)^{4}\notag\\
&\hspace{1.8cm}+ 20 z_{c}^{2}(z_{v}-1)^{5} + 8 z_{c}(z_{v}-1)^{6} + (z_{v}-1)^{7}\Bigr)\,(z_v-1)\,(z_c+z_v-1)=0.
\end{align}
Equivalently, $z_v+a\,z_c=1$ with
$a\in\left\{-1,\frac{1-\sqrt5}{2},0,\frac{3-\sqrt5}{2},1,\frac{1+\sqrt5}{2},2,\frac{3+\sqrt5}{2},3\right\}$,
and the smallest temperature is again at $a=3$, i.e.\ $z_v+3z_c=1$, which again coincides with the Hagedorn temperature of $\mathcal N=4$ SYM.

\section{Density of the Tower of Light Higher-spin States} \label{App: HS Density}

One of the most relevant properties of a tower of states is its density. In fact, it typically determines the origin of the tower in string compactifications. For instance, it determines the number of extra dimensions that become large at the same rate in decompactification limits. Similarly, one might hope that this quantity reflects the type of the leading string in tensionless string limits in AdS/CFT. Unlike the Hagedorn temperature, which, as discussed in Section~\ref{ss:alpha-vs-T_H}, receives contributions from any string whose tension is of the order of the AdS scale, the density of the lightest HS tower is more naturally related only to the leading tensionless string. With this motivation in mind, in this appendix we consider the density of states for the light HS tower that appears in the weak-coupling limit of gauge theories. We compute this density of states explicitly for $\mathcal{N}=4$ SYM, finding that the tower of light HS modes in the bulk is less dense than any known tower in string compactifications to flat space. Finally, we point out that this density of light HS states has been argued to hold for more general gauge theories in the large-spin limit. Hence, it does not seem to serve as a good way to distinguish different types of tensionless strings in AdS.

\medskip

At the free point of a gauge theory, HS conserved currents acquire an anomalous dimension at second order in the gauge coupling. In particular (see e.g.~\cite{Perlmutter:2020buo}),
\begin{equation}
    \gamma_J \sim f(J)\, g^2 \, \label{gamma},
\end{equation}
with $f(J)$ an analytic function of the spin $J$. The overall factor of $g^2$ is what we relate to the exponential decay appearing in the CFT Distance Conjecture. We now argue that the density of the tower in the bulk is determined by the large-spin behavior of $f(J)$.

The equation above ignores possible degeneracies at fixed spin, which could in principle be relevant for counting the density of states. However, these degeneracies will not affect the \emph{scaling} of the density of states with $J$. In the free theory, HS conserved currents are built by acting with $J$ derivatives on a bilinear of free fields. Hence, the number of HS conserved currents per spin is determined by (1) the number of bilinears that one can construct and (2) the number of HS conserved currents of spin $J$ that one can generate by acting with $J$ derivatives on a given bilinear. The first item is fixed by the number of fields and their representation under the gauge group. Since it will not affect the scaling with $J$, we can safely ignore this factor. As for the second one, it turns out that there is exactly one spin-$J$ current per bilinear. This can be seen by constructing the most general ansatz for a linear combination of $J$ derivatives acting on a bilinear of free bosons, fermions, or vectors and imposing that the result is a traceless, and conserved current (see e.g.~\cite{Skvortsov:2015pea}). All in all, we conclude that the density of the HS tower is determined by $f(J)$, up to an overall $J$-independent degeneracy factor at fixed spin, given by $n_{\rm bil}$.

So far we have discussed the density of the tower of operators in the CFT. Due to the non-linear AdS/CFT relation between conformal dimensions and bulk masses, this is not equivalent to the density of states in the bulk. For spin-$J$ traceless symmetric representations, we have
\begin{equation}
    m_J^2 L^2 = (\Delta + J - 2)(\Delta - J - d + 2)
              = (\gamma + 2J + d - 4)\,\gamma \,,\label{m}
\end{equation}
where $L$ is the AdS radius and we have introduced the anomalous dimension
\(\gamma = \Delta - J - d + 2 \, \).
For $\gamma \ll 1$ and $J \gg 1$, from \eqref{gamma} and \eqref{m} we obtain
\begin{equation}
    m_J L \sim \sqrt{2 J f(J)}\, g \, .
\end{equation}
Treating $L$ as fixed, we then see that the density of the HS tower at large $J$ behaves as
\begin{equation}
    \rho_J \sim \sqrt{J f(J)} \, .
\end{equation}

As discussed above, to fix the density of the tower, we only need to determine the large-$J$ behavior of $f(J)$. For concreteness, let us focus on $\mathcal{N}=4$ SYM, for which the anomalous dimensions of the HS currents on the leading single-trace Regge trajectory obey \cite{Anselmi:1998ms,Dolan:2001tt,Kotikov:2001sc}
\begin{equation}
    \gamma_J
      = \frac{g_{\rm YM}^2 N}{2 \pi^2}\, H(J)
        + \mathcal{O}(g_{\rm YM}^4) \,,
    \qquad
    H(J) = \sum_{n=1}^J \frac{1}{n} \, .
\end{equation}
For large spin, and keeping $N$ (or equivalently $L$ in Planck units) fixed, we have
\begin{equation}
    f(J) \sim H(J) \sim \log J \, .
\end{equation}
Thus, the density of the light HS tower is
\begin{equation}
    \rho_J \sim \sqrt{J \log J} \, .
\end{equation}
This should be contrasted with the density of light states in a tensionless-string limit in flat space, where one finds an exponential density of states. In fact, this density of states is even weaker than that of the Kaluza--Klein (KK) tower associated with a single extra dimension, which is the least dense tower known in flat-space string theory.

Let us conclude by noting that this large-$J$ behavior of $f(J)$ has been argued to hold for general weakly coupled gauge theories. This type of $\log J$ behavior was discussed extensively in \cite{Belitsky:2006en} and is intimately related to the existence of massless spin-one particles in the spectrum. Hence, the density of states of the light HS tower in tensionless string limits in AdS/CFT appears to be completely universal. Despite being an interesting feature of these limits, this implies that the density of the HS tower cannot be used to determine which type of string becomes tensionless at the fastest rate.

\bibliographystyle{JHEP}
\bibliography{references}
\end{document}